\documentclass[table,xcdraw]{article}
\usepackage{amsmath}
\usepackage{amssymb}
\usepackage[a4paper, total={6in, 8in},margin=1in]{geometry}
\usepackage{graphicx}
\usepackage{wrapfig}
\usepackage{calrsfs}
\usepackage[utf8]{inputenc}
\usepackage{fourier} 
\usepackage{array}
\usepackage{makecell}
\usepackage{hyperref}
\usepackage{cite}
\usepackage{multirow}
\usepackage{mathrsfs}
\usepackage{titlesec}
\titleformat{\section}{\Huge}{\thesection}{1em}{}
\usepackage{slashed}
\usepackage{caption}
\usepackage{comment}
\usepackage{placeins}
\usepackage{tikz-feynman,contour}
\usepackage{xcolor}

\DeclareMathAlphabet{\pazocal}{OMS}{zplm}{m}{n}
\newcommand{\nobs}{$n_{\text{obs}}$}

\begin{document}

\title{Mono-$X$ Signal Characterization from Two-component \\[0.25cm]
Dark Matter Using a Convolutional Neural Network}
\author{Max Fusté Costa$^{a,b}$, Yong Sheng Koay$^a$ and Stefano Moretti$^{a,c}$}
\maketitle
\centerline{\large $^a$Department of Physics \& Astronomy, Uppsala University, Box 516, 75120 Uppsala, Sweden}
\vspace{0.5cm}
\centerline{\large $^b$Deutsches Elektronen-Synchrotron DESY, Notkestr. 85, 22607 Hamburg, Germany}
\vspace{0.5cm}
\centerline{\large $^c$School of Physics \& Astronomy, University of Southampton, Southampton SO17 1BJ, UK}
\vspace{2.0cm}
\centerline{\Large Abstract}
\vspace{1.0cm}
\noindent
We assess the  scope of a Convolutional Neural Network (CNN) in characterizing potential signals of two-component Dark Matter (DM) arising at the Large Hadron Collider (LHC) from mono-jet and mono-$Z$ probes. We show that such a CNN has the ability of not only  inferring the presence of two DM particles but also of  extracting their mass and spin, the latter being either 0 or 1/2, following detector level analysis. However, such result represents a conceptual proof-of-concept, as we have not entertained a signal-to-background analysis.  
\vspace{9cm}
\hrule
\vspace{0.25cm}\noindent
{Emails:\\ {max.fuste.costa@desy.de; yongsheng.koay@physics.uu.se; stefano.moretti@cern.ch; harri.waltari@physics.uu.se}}
\newpage
\tableofcontents
\newpage

\pagenumbering{arabic}
\section{Introduction}\label{sec:Intro}

Dark Matter (DM), a hypothetical form of matter that does not interact with Electro-Magnetic (EM) or strong radiation in any way, still remains one of the most prevalent unsolved questions in cosmology as well as particle physics and the main evidence for its existence come from the gravitational effects that it has on baryonic matter \cite{Zwicky:1933gu,Rubin:1970zza,Corbelli:1999af,Clowe:2006eq,Planck:2018vyg,Bertone:2004pz}. 

\noindent Despite years of collective efforts using different search methods, DM has not yet been detected. Therefore, its exact nature is unknown. It is thought to be non-baryonic and most likely composed of some subatomic particles that have yet to be discovered. Even though DM is often assumed to be a single particle (i.e., one-component DM), there is no reason to rule out the possibility of a dark sector similar to the one for ordinary matter, which is described by the Standard Model (SM) of particle physics, which has multiple (similar) copies of both quarks and leptons. In this paper, to realize such a scenario, a minimal extension of the SM is studied, utilizing two DM particles, a fermion and a scalar (i.e., two-component DM), together with a scalar mediator in both cases. 

\noindent One of the three main paths of research to detect DM particles is to produce these in high energy particle accelerators \cite{Buchmueller:2017qhf}, like the Large Hadron Collider (LHC). Herein, collision events sometimes produce certain invisible particles that escape the detector due to these not interacting via the EM or strong forces. The Missing Transverse Energy (MET or $E_T^{\rm miss}$) in the detector, therefore, is used as a signature for their presence. Accurate measurements of this quantity require full and precise reconstruction of all visibly produced particles in the interaction, in the quest to separate DM signals from the SM background (comprising of neutrinos).

\noindent In the case of DM being detected at a high energy collider, the signal would be analyzed to determine the number of distinct particles that it contained, along with their most relevant characteristics, such as the mass and the spin. The analysis and further characterization of the data is notoriously complex for a human, hence promoting the use of  Neural Networks (NNs). In fact, in the last decade, the latter have become one of the most popular methods of Machine Learning (ML) applied to fundamental sciences. In physics, they have been used to solve a myriad of problems, both in classical and quantum physics \cite{Dunjko:2018xgc,Krenn:2016srv,Koch-Janusz:2017jhf}, including particle physics \cite{Feickert:2021ajf}.

\noindent We shall study the DM signatures through the use of Monte Carlo (MC)  simulations. Specifically, in this study, these methods are used to generate some of the most common DM  signals at the LHC, the so-called mono-$X$ signatures, wherein a single object described by the SM is produced alongside a pair of invisible DM particles \cite{Brennan:2016xjh,Liew:2016oon,Bernreuther:2018nat}. In particular, the considered signals  produce either a single jet or a $Z$ boson, which give name to their respective signatures, mono-jet and mono-$Z$. Obviously, if such signals were seen at the LHC, one would wish to do  cross checks in other SM channels therein. Furthermore, one might also like to know how wide a range of masses do we need to study in, say, non-collider experiments seeking (in)direct detection of DM.

%Given that we are working only with simulated data, this put us in the domain of simulation-based inference \cite{Brehmer:2020cvb,Cranmer:2019eaq}.\KYS{This is not sbi. sbi requires pdfs. rewrite needed.} 
\noindent There exist previous studies \cite{Khosa:2019kxd,Arganda:2021azw} (see also 
\cite{Celik:2023dis,Lv:2022pme,Finke:2022lsu})  on inferring DM models using ML methods (NNs or others). However, they have predominantly focused on one-component DM scenarios. Furthermore, have been  applied on specific benchmark points in parameter space and have primarily employed the mono-jet signature as the experimental observable. In this work, we extend these efforts by considering two-component DM models, applying ML on the relevant parameter space and also incorporating the mono-$Z$ signature. Our approach not only aims at inferring the underlying DM model but also seeks to determine the properties pertaining to the corresponding DM candidate(s), like mass and spin. The presence of a second DM component also brings in a new  kind of background to the problem, when compared to the case of one-component DM  (as we shall discuss below). 

\noindent The generated DM signals, all at Leading Order (LO), contain either one- (a fermion or a scalar) or two-component DM  data (with the two produced simultaneously). We anticipate here that, with the information provided by the more numerous mono-jet signature, the NN is expected to distinguish the number of DM components in the signal  and then determine their masses. With the less numerous mono-$Z$ signature,  the NN is expected to exploit such an acquired knowledge to identify whether an individual DM signal component  corresponds to a fermion or a scalar. In order to achieve this, a Convolutional NN (CNN) algorithm is exploited. However, we should  state from the outset that, in the present paper, we solely intend to explore the potential of ML in characterization tasks of these two mono-$X$ signatures under the hypothesis of a background free environment, which is clearly ideal in most experimental conditions, yet, we believe that this, so to say,  `proof-of-concept' is interesting in the general endeavor of understanding the nature of DM once detected at colliders, chiefly, the LHC and its High-Luminosity version (HL-LHC)  \cite{Gianotti:2002xx}. 

\noindent The plan of the paper is as follows. We start with some theoretical background, wherein the aforementioned DM model is introduced, along with the used mono-$X$ signatures and the information on the DM that can be extracted from each of these. In the following section, the framework used to simulate the collision events that produce the two DM particles is described, alongside the implementation of the CNN for each individual DM candidate. Then, there is a discussion of the CNN performance and an assessment of the quality of the information extracted. Finally, we conclude.

\section{DM Components}\label{sec:model}

DM needs to be stable on cosmological timescales. The usual way of achieving this is to extend the particle content of the SM and impose some new symmetry onto the extension, making the lightest additional   particle charged under this symmetry stable. To achieve two DM candidates, one usually needs to implement two such symmetries, although sometimes  one symmetry can stabilize two particles, if the heavier has no allowed decay modes (or only allowed over a timescale longer than the universe age). There exists a large number of two-component DM models in the literature \cite{Bhattacharya:2013hva,Esch:2014jpa,Aoki:2016glu,PeymanZakeri:2018zaa,Bernal:2018aon,Borah:2019aeq,Belanger:2020hyh,Khalil:2020syr,Khalil:2021tpz,Chakrabarty:2021kmr,Yaguna:2021rds,Aranda:2019vda,Hernandez-Otero:2022dxd,Hernandez-Sanchez:2020aop,Hernandez-Sanchez:2022dnn} and some with more than two candidates \cite{Aoki:2012ub,Yaguna:2019cvp,Belanger:2022esk}. Here, as theoretical framework, we have chosen the  model of Ref.~\cite{Esch:2014jpa}, where one extends the SM with two DM candidates, one fermion and one scalar, plus a scalar mediator, which all are singlets under the SM gauge group ${\rm SU}(3)_C\times {\rm SU}(2)_L \times {\rm U}(1)_Y$.

\noindent The Lagrangian terms involving the DM fermion ($\chi$) are written as follows:
\begin{equation}
        \mathcal{L} = - \frac{1}{2} (M_{\chi} \bar{\chi} \chi + g_S \phi \bar{\chi} \chi + g_P \phi \bar{\chi} \gamma_5 \chi), \label{eq:fermlagrangian}
\end{equation}
\noindent where $M_{\chi}$ is the mass of the DM fermion while  $g_S$ and $g_P$ are the scalar (CP-even) and pseudoscalar (CP-odd) couplings of $\chi$ to the spin-0 current.

\noindent The scalar potential of this model is given by: 
\begin{eqnarray}
    V(\phi,H,S) & = & - \mu^2_h (H^\dagger H) + \lambda_H (H^\dagger H)^2 + \frac{\mu^2_{\phi}}{2} \phi^2 + \frac{\lambda_{\phi}}{4}\phi^4 + \mu^3_1 \phi
     + \frac{\mu_3}{3} \phi^3 + \frac{\lambda_4}{2} \phi^2 (H^\dagger H) \nonumber \\ 
     & &+ \mu \phi (H^\dagger H) + \frac{1}{2} \mu^2_S S^2 + \frac{\lambda_S}{4} S^4
     + \frac{\lambda}{4} (H^\dagger H) S^2 + \frac{\mu_{\phi S S}}{2} \phi S^2 + \frac{\lambda_{\phi \phi S S}}{2} \phi^2 S^2,\label{eq:scalarpotential}
\end{eqnarray}
\noindent where $H$ is the SM Higgs doublet, $\phi$ is the mediator scalar and $S$ the DM scalar. Both $\phi$ and $S$ are real fields.

\noindent If we limit ourselves to dimension-4 operators, one needs to impose only one additional $Z_2$ symmetry. The SM particles and the mediator scalar are even under such a symmetry while both DM candidates are odd. The heavier DM candidate is also stable as one cannot write an interaction term involving both $\chi$ and $S$, thereby yielding a two-component DM model.

\noindent The real part of the neutral component of the Higgs doublet, $h$, and the mediator field, $\phi$, mix with each other as a result of the $\mu$ term, giving rise to two mass eigenstates, $H_1$ and $H_2$, which are defined as
\begin{equation}\label{eq:mixing}
    H_1 = h \cos(\alpha) + \phi \sin(\alpha), \qquad H_2 = \phi \cos(\alpha) - h \sin(\alpha), 
\end{equation}
\noindent where $\alpha$ is the mixing angle, which is assumed to be small. Thus $H_1$ can be identified with the observed Higgs boson with a mass of 125 GeV. 
(Here, $H_{1}$ can be either the lighter or the heavier one of the two mass eigenstates.) As long as the mixing is sufficiently small, below $\sim 4\%$, the experimental constraints of Ref.~\cite{LEPWorkingGroupforHiggsbosonsearches:2003ing} from low mass Higgs searches can be satisfied. For masses above the $W$ mass the constraints are much  weaker. For such a case the strongest constraints arise from the signal strengths of the discovered 125 GeV Higgs boson \cite{ATLAS:2019nkf,CMS:2022dwd}, which allow a mixing of $\sim 10\%$.

\begin{figure}[!ht]
\begin{center}
\begin{tikzpicture}
\begin{feynman}
      \node[blob] (m) at ( 0, 0);
      \vertex (a) at (-1.5,-1.5) {p};
      \vertex (b) at ( 1.5,-1.5) {DM};
      \vertex (c) at (-1.5, 1.5) {p};
      \vertex (d) at ( 1.5, 1.5) {DM};
      \diagram* {
        (a) -- (m) -- (c),
        (b) -- (m) -- (d),
      };
      \vertex (r) at ( 0, 1.75);
      \draw [gluon] ($(c)!0.4!(m)$) -- (r);
\end{feynman}
\end{tikzpicture}
\end{center}
\caption{Example of topology giving rise to a mono-jet signature in a proton-proton collision into a DM pair, with a gluon jet arising from the initial state.\label{fig:monojet}}
\end{figure}
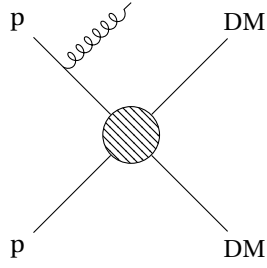

\noindent In a hadron collider like the LHC, mono-jet events, shown in Figure \ref{fig:monojet}, are most copious.
\noindent Furthermore, as the jet originates from an initial state gluon or quark and not from the DM particle itself, 
we expect that the mono-jet signature could give information only on the mass of the DM particles, as the jet recoils against the DM pair and therefore its  transverse momentum, $p_T^j$, or, similarly, the MET, distribution carries imprints of it. To some extent, also the pseudorapidity, $\eta$,  does. 

\begin{figure}
\begin{center}
\begin{tikzpicture}
\begin{feynman}
      \node[blob] (m) at ( 0, 0);
      \vertex (a) at (-1.5,-1.5) {p};
      \vertex (b) at ( 1.5, 0) {DM};
      \vertex (c) at (-1.5, 1.5) {p};
      \vertex (d) at ( 1.5, 1.5) {DM};
      \vertex (z) at ( 1.5, -1.5);
      \vertex (f2) at ( 2.5, -1) {l+};
      \vertex (f1) at ( 2.5, -2) {l-};
      \diagram* {
        (a) -- (m) -- (c),
        (b) -- (m) -- (d),
        (m) -- [boson, edge label=\(Z\)](z),
        (z) -- [fermion] (f1),
        (z) -- [anti fermion] (f2),
      };
\end{feynman}
\end{tikzpicture}
\end{center}
\caption{Generic topology giving rise to a mono-$Z$ signature in a proton-proton collision into a DM pair, with a $Z$ boson (decaying into a lepton pair, $l=e,\mu$) arising from (potentially) anywhere in the event.\label{fig:monoZ}}
\end{figure}
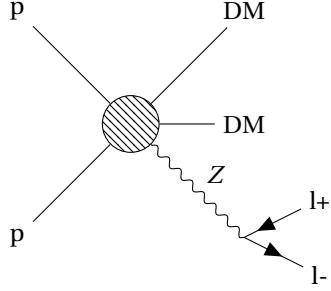

\noindent Other mono-$X$ signatures could carry further information on the DM candidates, though, like mono-$Z$,
shown in Figure \ref{fig:monoZ}. For instance,  angular correlations between the $Z$ decay products (e.g., an electron and/or muon) and the direction of the $E_T^{\rm miss}$
could be sensitive to the DM spin as, unlike a jet, the $Z$ boson can also be produced through final state radiation \cite{Abdallah:2019tpo}. In order to achieve this, we need to make the DM candidates charged under the SU($2)_L$ group. The simplest way to do so is to promote the DM candidates to be SU$(2)_L$ doublets:
\begin{align}
    \chi =
    \begin{pmatrix}
    \chi^+ \\
    \chi^0
    \end{pmatrix}, \qquad
    S =
    \begin{pmatrix}
    S^+ \\
    S^0
    \end{pmatrix}.
    \label{eq:doublet}
\end{align}

\noindent The scalar potential of equation (\ref{eq:scalarpotential}) will  then be the same with only the modification $S^{2}\rightarrow S^{\dagger}S$, while the fermionic part (\ref{eq:fermlagrangian}) needs to be modified with the proper SU$(2)_L$ contractions.

\noindent Similarly to the mono-jet signature, it is possible to gather (further) information on the mass of the DM particles through the mono-$Z$ signature too, which means that the $p_T$ and $\eta$ of either of the outgoing leptons and the MET are of interest here. Additionally, since the $Z$ boson can couple to the final state (as mentioned), it is possible to obtain information on the spin of the DM particle(s) produced, e.g., {through the difference of the angular distances $\phi$ between each of the leptons and the MET in the transverse plane}. Furthermore, although the event rate for the mono-$Z$ process is smaller than for mono-jet, the final state with leptons is a clean one and would have smaller backgrounds.

\noindent Once the DM candidates are promoted to SU$(2)_{L}$ doublets, the $Z$ boson becomes a possible mediator. In practice, the gauge coupling is so large  that it becomes the dominant mediator and the singlet scalar $\phi$ can be neglected. Since there is no tree-level $ZZZ$ vertex in the SM, the $Z$ boson cannot be radiated from the mediator.

\section{Training Data}\label{sec:data}

We implemented our two-component DM model described in section \ref{sec:model} in \textsc{SARAH} v4.14.5 \cite{Staub:2008uz,Staub:2012pb,Staub:2013tta}. As \textsc{SARAH} does not support tadpole terms, we omitted the tadpole term from equation (\ref{eq:scalarpotential}) and allowed the $\phi$ field to have a Vacuum Expectation Value (VEV) instead. We generated the spectra and couplings with \textsc{SPheno} v4.0.5 \cite{Porod:2003um,Porod:2011nf}. The simulations are performed with $\textsc{MadGraph5\_aMC@NLO}$ \cite{Alwall:2014hca}, using the packages $\textsc{Pythia8}$ \cite{Sjostrand:2007gs,Sjostrand:2014zea} for parton showering/hadronization and $\textsc{Delphes3}$ \cite{deFavereau:2013fsa} for detector simulation, while the event distributions are analyzed with $\textsc{MadAnalysis5}$ \cite{Conte:2012fm}.

\noindent The model, introduced in \cite{Esch:2014jpa}, is used to simulate proton-proton collisions that produce the two DM particles. This is studied through two signatures, mono-jet and mono-$Z$, described in equations (\ref{eq:monojet}) and (\ref{eq:monoZ}), respectively, and Figures \ref{fig:monojet} and \ref{fig:monoZ}, found in the previous section. As mentioned therein, for the mono-jet case we use the model with singlet DM candidates and for the mono-$Z$ signature the one with doublet DM candidates to allow final state radiation.

\noindent The LHC processes that we target are (here, $j=$~jet)
\begin{equation}\label{eq:monojet}
    p + p \rightarrow \chi + \bar{\chi} + j, \qquad p + p \rightarrow S + S + j,
\end{equation}    
\begin{equation}\label{eq:monoZ}
    p + p \rightarrow \chi + \bar{\chi} + Z, \qquad p + p \rightarrow S + S + Z,
\end{equation}
with the $Z$ boson decaying into a pair of leptons, either an electron/positron or a muon/anti-muon pair.

\noindent We now discuss the data generation for the two signatures in turn.

\subsection{Data Generation}
\subsubsection{Mono-Jet}

\noindent For simplicity we set the pseudoscalar coupling of $\chi$, $g_P$, to zero and thus the only coupling between the mediator scalar and the fermion is the scalar one. Thus, equation (\ref{eq:fermlagrangian}) is rewritten as
\begin{equation}\label{eq:fermlagrangian-simplified}
    \mathscr{L} = - \frac{1}{2} (M_{\chi} \bar{\chi} \chi + g_S \phi \bar{\chi}\chi).
\end{equation}

\noindent In this model, just like in other multi-component DM models, the cross sections of the two DM particles are not necessarily of the same order of magnitude, as they depend on both the couplings and the masses of the particles. We take as a part of the definition of a two-component DM scenario that both components should contribute sizably to the relic density. This in turn means that the cross sections of the processes $\mathrm{DM}+\mathrm{DM}\rightleftarrows \mathrm{SM}+\mathrm{SM}$ should be of similar order (outside possible resonances) if the masses are comparable. One of these processes determines the annihilations in the early Universe, the other determines the chances of producing DM at colliders.

\noindent This requirement also affects the ML problem. In principle, both of the components should be detectable from the signal, which would not be the case if one of the cross sections was several orders of magnitude larger than the other. {{The main part of this study works with the case when the cross sections are similar (as defined in equation (\ref{eq:XS_ratios} below) for the case where $r=1$): }}
%
%However, we also} vary the ratio of the cross sections between the two-components and see how this affects the {\textcolor{orange}{ ability of the model to predict the masses of both components. We explore the cases where the cross section of the fermionic component has a cross section that is $r=2$, $3$ and $5$ times bigger than that of the scalar component,}}  as it has the larger spin multiplicity in the final state. 
%
\begin{equation}\label{eq:XS_ratios}
    0.85<\frac{\sigma_{fermion}}{(r\cdot\sigma_{scalar})}<1.15.
\end{equation}

\noindent The cross section depends on both the mass and the couplings between the DM particles and either the Higgs boson or the $\phi$ scalar mediator, which mix (as discussed). We try to aim at a relatively constant cross section by varying the couplings as a function of the mass. With such scaling of the couplings one may ensure that the cross sections for the two DM particles are in a given ratio. We vary the masses within the range $m \in [10,500]$~GeV for both particles, which is a reasonable range for the LHC experiments. 

\subsubsection*{Constraints on the Couplings}

\begin{figure}
\begin{center}
\begin{tikzpicture}
\begin{feynman}
    \vertex (a) {\(\phi\)};
    \vertex [right=of a] (b);
    \vertex [above right=of b] (f1) {$\chi$};
    \vertex [below right=of b] (f2) {$\bar{\chi}$};

    \diagram* {
      (a) -- [scalar] (b) -- [fermion] (f1),
      (b) -- [anti fermion] (f2)
    };
\end{feynman}
\end{tikzpicture}
\end{center}
\caption{Coupling between the DM fermion $\chi$ and the scalar mediator $\phi$. The coupling constant related to this interaction is $g_S$.\label{fig:PhiChiChi}}
\end{figure}
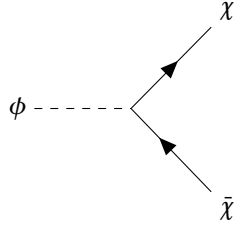

\noindent By taking a look at equation (\ref{eq:fermlagrangian-simplified}), we can see that the only coupling of the DM fermion, $\chi$, and the mediator is via the vertex in Figure~\ref{fig:PhiChiChi}, which is given by the coupling $g_S$. This vertex is distributed to the two mass eigenstates $H_{1}$ and $H_{2}$, the mostly-singlet state $H_{2}$ being the more important one as the $H_{1}\chi\tilde{\chi}$ coupling is suppressed by the small mixing angle $\alpha$ of equation (\ref{eq:mixing}).

\noindent In order to get scenarios with two DM components produced in roughly equal amounts, we choose the value of the coupling $g_S$ so that the cross section has a nearly constant value for all masses. This means that for higher masses the coupling is larger. We start at the highest mass in our range and pick a large value for the coupling. The coupling cannot be arbitrarily large or the theory becomes non-perturbative. As the perturbative limit we choose
\begin{equation}
    g_S < 3. 
\end{equation}

\noindent As we will need the largest value of $g_{S}$ at the highest masses and the cross section scales as $\sigma \propto g_{S}^{2}$, we can get a constant cross section for the process if we choose the couplings according to
\begin{equation}
    g_S (m_{\chi}) = g_S (m_{\chi {\rm max}}) \sqrt{\frac{\sigma(m_{\chi})}{\sigma(m_{\chi {\rm max}})}},\label{eq:scaling}
\end{equation}
\noindent where the cross sections $\sigma(m_{\chi})$ are initially computed with a fixed coupling value.

\begin{figure}[!ht]
    \centering
    \includegraphics[width=0.48\linewidth]{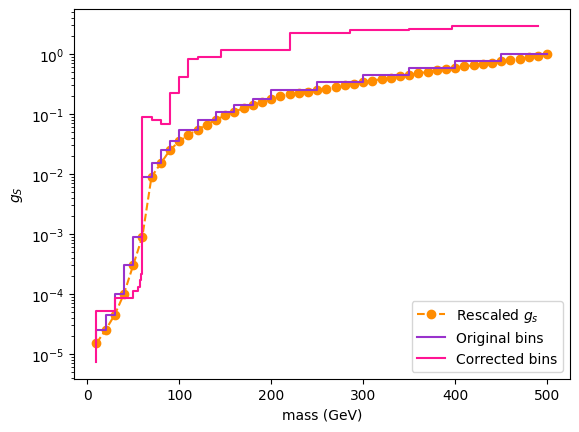 }
    \hfill
    \includegraphics[width=0.48\linewidth]{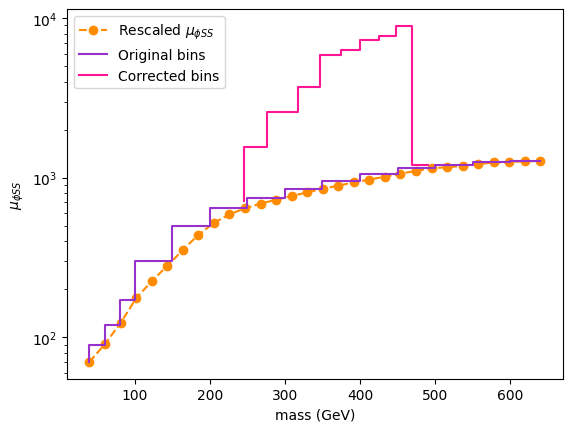}
    \caption{Rescaling of the coupling constants for the mono-jet case. The orange curve, dotted, represents the rescaled values of the constants after the aforementioned process. The purple line, continuous, draws the binning of the orange curve, while the pink line, also continuous, draws the corrected binning of the orange curve that ensures that the cross section is of a similar order of magnitude for both processes. The sudden change in the value of the coupling constant is caused by the mass of the mediator field, $\phi$, and the one in the corrected binning of $\mu_{\phi S S}$ is caused by $\lambda_{\phi \phi S S}$ and $\lambda$ as well as their associated potentials. Also, it is relevant to note that, in the right plot, the bin that encompasses the masses between 100 and 200 GeV was suppressed from the plot for visibility, as the coupling constant drops to 2.64.}
    \label{fig:rescale}
\end{figure}

\noindent This rescaling gives knowledge on the shape of the curve of the coupling constant as a function of the $\chi$ mass, as can be seen in Figure \ref{fig:rescale} (left). For $m_{\chi}\ll m_{\phi}$ and $m_{\chi}\gg m_{\phi}$, the coupling varies quite smoothly, whereas around $m_{\chi}\simeq m_{\phi}$ the size of the coupling changes by several orders of magnitude in a short mass interval due to DM production shifting from on-shell $\phi$ decays to production through an off-shell (virtual) $\phi$ boson. We now choose mass intervals, where the coupling is nearly constant and give the coupling just one value throughout the interval.

\begin{figure}
\begin{center}
\begin{tikzpicture}
\begin{feynman}
    \vertex (a) {\(\phi\)};
    \vertex [right=of a] (b);
    \vertex [above right=of b] (f1) {$S$};
    \vertex [below right=of b] (f2) {$S$};

    \diagram* {
      (a) -- [scalar] (b) -- [scalar] (f1),
      (b) -- [scalar] (f2)
    };
\end{feynman}
\end{tikzpicture} \qquad
\begin{tikzpicture}
\begin{feynman}
    \vertex (m);
    \vertex [above left=of m] (a) {\(\phi\)};
    \vertex [below left=of m] (b) {\(\phi\)};
    \vertex [above right=of m] (f1) {$S$};
    \vertex [below right=of m] (f2) {$S$};

    \diagram* {
      (a) -- [scalar] (m) -- [scalar] (f2),
      (b) -- [scalar] (m) -- [scalar] (f1)
    };
\end{feynman}
\end{tikzpicture} \qquad
\begin{tikzpicture}
\begin{feynman}
    \vertex (e);
    \vertex [above left=of e] (c);
    \vertex [below left=of e] (d);
    \vertex [left=of c] (a);
    \vertex [left=of d] (b);
    \vertex [right=of e] (f);
    \vertex [above right=of f] (g) {$S$};
    \vertex [below right=of f] (h) {$S$};
    \vertex (i) {\(H\)};
    
    \diagram* {
      %(a) -- [gluon] (c) -- 
      (e) -- (i) --
      [scalar, edge label=\(\)] 
      (f) -- 
      [scalar] (g),
%      (b) -- [gluon] (d) -- (e),
      (c) -- (d),
      (f) -- [scalar] (h)
    };
\end{feynman}
\end{tikzpicture} 
\end{center} 
\caption{Couplings between the DM scalar $S$ and the mediator $\phi$ (left and centre) as well as the SM Higgs state (right). The associated constants are, from left to right, $\mu_{\phi S S}$, $\lambda_{\phi \phi S S}$ and $\lambda$. Once $\phi$ acquires a VEV, the second diagram also contributes to the three-point interaction $\phi S S$.\label{fig:PhiSS}}
\end{figure}
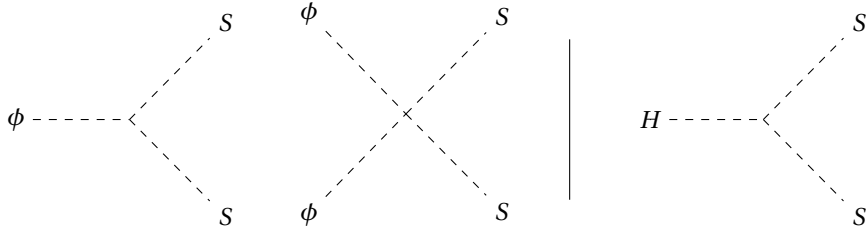

\noindent A similar procedure can be adopted for the DM scalar, although it is not as simple. In fact, in the process (\ref{eq:monojet}), $S$ couples to the mediator through three different couplings (see Figure~\ref{fig:PhiSS}). Here both $H_{1}$ and $H_{2}$ can have unsuppressed couplings and hence a simple scaling relation like equation (\ref{eq:scaling}) cannot be written.

\noindent There are altogether three parameters that are related to the production cross section of scalar DM, arising from vertices shown in Figure \ref{fig:PhiSS}. We choose $\langle \phi \rangle$ to be small so that it will not contribute largely to the SM Higgs properties through their interaction terms in equation (\ref{eq:scalarpotential}). As the first diagram, apart from a mass dimension,  is similar to the one in the fermionic DM case, we choose to mainly vary $\mu_{\phi SS}$. We also set a bound for perturbativity for this coupling similar to the one for $g_S$: 
\begin{equation}
    \mu_{\phi S S} < 3 m_{\phi}, 
\end{equation}
where $m_{\phi}$ is the mass of the mediator. This constant can be adjusted to give a cross section comparable to the fermionic DM production with an analogous process. However, since the two coupling constants that contribute to the cross section will not be altered and the cross sections of both DM particles are required to be of a similar order of magnitude, this process is not trivial and will require some adjusting of $g_S$. 

\noindent {{In order to ensure that all the aforementioned conditions are met, the available mass space is to be reduced, carefully picking pairs of masses that result in cross sections that fulfill equation \ref{eq:XS_ratios}. Therefore, for a given mass of the fermionic component, the corresponding value of $g_S$ must be picked as given by Figure \ref{fig:rescale}. This mass must be paired with a mass of the scalar component that is within the available range given by Figure \ref{fig:mass_space_jet}, which also has a corresponding value of $\mu_{\phi s s}$. The phase space for the cases where there is only $\chi$ or $S$ in the final state is the same as the one where both particles are present.} 

\begin{figure}[!ht]
    \centering
    \includegraphics[width=0.7\linewidth]{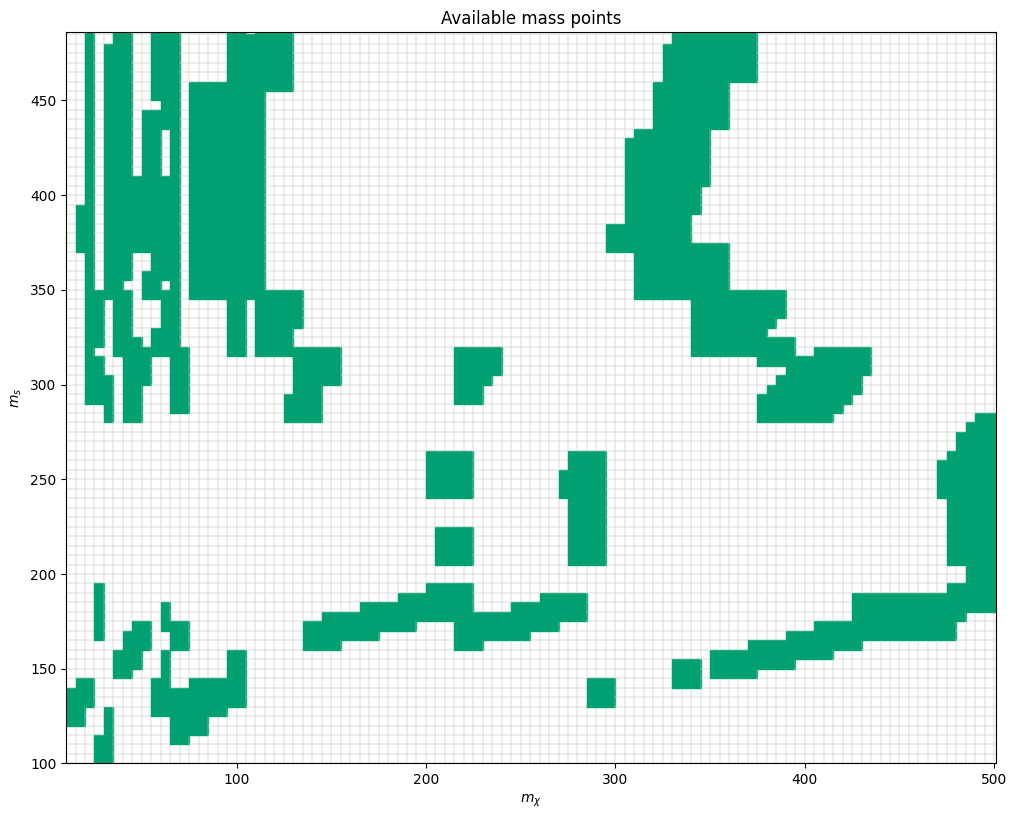 }
    \caption{Phase space of the masses of the fermionic and scalar component that fulfill all the given conditions. The overall mass range for the fermionic component is  $m_{\chi} \in [10,500]$ while for the scalar we have $m_S \in [100,485]$.}
    \label{fig:mass_space_jet}
\end{figure}

\subsubsection{Mono-$Z$}

In the mono-jet case, both $\chi$ and $S$ are SU$(2)_L$ singlets that do not couple to gauge bosons so $\phi$ is the only possible mediator. In the mono-$Z$ case, however, the two DM candidates need to be able to couple to the $Z$ boson, as the only mono-$Z$ producing channels involve a coupling between a quark-antiquark pair and a DM pair through a $Z$ boson mediator, as shown in Figure~\ref{fig:monoZ-New}.

\begin{figure}[!t]
\begin{center}
\begin{tikzpicture}
\begin{feynman}
    \vertex (c);
    \vertex [above left=of c] (a) {\(q\)};
    \vertex [below left=of c] (b) {\(\bar{q}\)};
    \vertex [right=of c] (d);
    \vertex [above right=of d] (e) {\(S\)};
    \vertex [below right=of d] (f) {\(S\)};

    \diagram* {
      (a) -- (c) -- [boson,edge label=\(Z\)] (d) -- [scalar] (e),
      (b) -- (c),
      (d) -- [scalar] (f),
    };
    \vertex (r) at ( 0, 1);
    \draw [boson] ($(a)!0.6!(c)$) -- (r);
\end{feynman}
\end{tikzpicture} \qquad
\begin{tikzpicture}
\begin{feynman}
    \vertex (c);
    \vertex [above left=of c] (a) {\(q\)};
    \vertex [below left=of c] (b) {\(\bar{q}\)};
    \vertex [right=of c] (d);
    \vertex [above right=of d] (e) {\(S\)};
    \vertex [below right=of d] (f) {\(S\)};

    \diagram* {
      (a) -- (c) -- [boson,edge label=\(Z\)] (d) -- [scalar] (e),
      (b) -- (c),
      (d) -- [scalar] (f),
    };
    \vertex (r) at ( 3, 0);
    \draw [boson] ($(d)!0.6!(f)$) -- (r);
\end{feynman}
\end{tikzpicture} \qquad
\begin{tikzpicture}
\begin{feynman}
    \vertex (c);
    \vertex [above left=of c] (a) {\(q\)};
    \vertex [below left=of c] (b) {\(\bar{q}\)};
    \vertex [right=of c] (d);
    \vertex [above right=of d] (e) {\(S\)};
    \vertex [below right=of d] (f) {\(S\)};
    \vertex (g) at (2.75,0.35);

    \diagram* {
      (a) -- (c) -- [boson,edge label=\(Z\)] (d) -- [scalar] (e),
      (b) -- (c),
      (d) -- [scalar] (f),
      (d) -- [boson] (g),
    };
\end{feynman}
\end{tikzpicture} \qquad
\begin{tikzpicture}
\begin{feynman}
    \vertex (c);
    \vertex [above left=of c] (a) {\(q\)};
    \vertex [below left=of c] (b) {\(\bar{q}\)};
    \vertex [right=of c] (d);
    \vertex [above right=of d] (e) {\(\chi\)};
    \vertex [below right=of d] (f) {\(\bar{\chi}\)};

    \diagram* {
      (a) -- (c) -- [boson,edge label=\(Z\)] (d) -- [fermion] (e),
      (b) -- (c),
      (d) -- [anti fermion] (f),
    };
    \vertex (r) at ( 0, 1);
    \draw [boson] ($(a)!0.6!(c)$) -- (r);
\end{feynman}
\end{tikzpicture} \qquad
\begin{tikzpicture}
\begin{feynman}
    \vertex (c);
    \vertex [above left=of c] (a) {\(q\)};
    \vertex [below left=of c] (b) {\(\bar{q}\)};
    \vertex [right=of c] (d);
    \vertex [above right=of d] (e) {\(\chi\)};
    \vertex [below right=of d] (f) {\(\bar{\chi}\)};

    \diagram* {
      (a) -- (c) -- [boson,edge label=\(Z\)] (d) -- [fermion] (e),
      (b) -- (c),
      (d) -- [anti fermion] (f),
    };
    \vertex (r) at ( 3.5, 1);
    \draw [boson] ($(d)!0.5!(e)$) -- (r);
\end{feynman}
\end{tikzpicture} 
\end{center} 
\caption{Channels that produce a pair of DM particles via a mono-$Z$ signature, with the $Z$  decaying into a pair of leptons, as depicted in Figure \ref{fig:monoZ}, wherein the mediator is the $Z$ boson itself. The top diagrams correspond to fermionic DM whereas the bottom diagrams correspond to scalar DM.}
\label{fig:monoZ-New}
\end{figure}
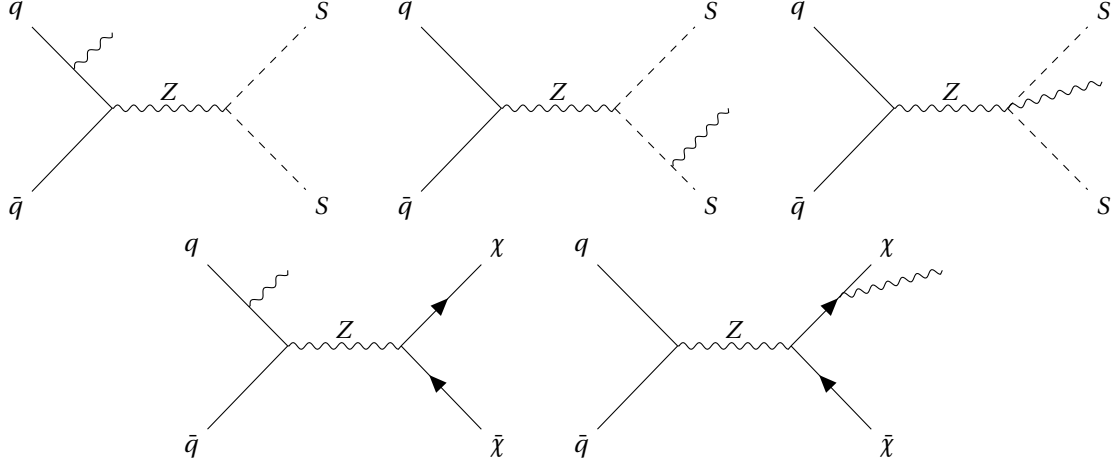

\noindent In order to make $\chi$ and $S$ couple to the $Z$ boson, they have been promoted to doublets under $SU(2)_L$, as described in equation (\ref{eq:doublet}), and partial derivatives have been promoted to covariant derivatives in the kinetic terms. The other changes, introduced in the previous section, are also implemented here, meaning that the mass and interaction terms for $\chi$ and $S$ are defined by equations (\ref{eq:scalarpotential}) and (\ref{eq:fermlagrangian-simplified}), respectively. 

\subsubsection*{Constraints}

Since the $Z$ boson can act as the mediator and the gauge coupling is relatively large, $g\simeq 0.6$, it easily becomes the dominant mediator compared to $\phi$. Hence, the production cross sections of DM particles depend practically only on their mass. Comparable cross sections can be obtained only if the masses of the DM candidates are relatively close to each other. In this case we set the couplings to $\phi$ as constants, $\mu_{\phi S S}=125$~GeV and $g_S=0.2$.

\noindent Since we want that the cross sections for the $\chi$ and $S$ cases are of a comparable order of magnitude also for mono-$Z$, this condition limits the distribution of the points in the mass space, since, for a certain value of $m_{\chi}$, only a certain range of values of $m_S$ is possible. To determine which pairs of masses have similar cross sections, each process is run separately and then ranges of both masses are paired together. We again consider a range of masses at the Electro-Weak (EW) scale leading to a sufficient number of signal events. 

%\begin{figure}[!ht]
%    \centering
%    \includegraphics[width=0.4\linewidth]{new_figs/allowed_masses_mono-Z.png }
%    \caption{Space of the masses of the fermionic and scalar component that fulfill all the given conditions. The mass ranges considered are $m_{\chi} \in [100,320]$~GeV and $m_S \in [115,500]$~GeV.\textcolor{red}{YS: if this is the space of mass to infer, it seems like knowing one automatically knowing the other in the 2-comp scenario, linear relation with some width. Good performance in mono-$Z$ might be because of this though. (either mention this upfront or state why mono-$Z$ is still better despite of this).  Also, this figuire was never mentioned in text.} }
%    \label{fig:mass_space_Z}
%\end{figure}

\subsubsection*{Mass Distributions}

{The preceding discussion on the constraints on the masses of the two DM components and their respective coupling constants refers to the Lagrangian mass terms in equations (\ref{eq:fermlagrangian}) and (\ref{eq:scalarpotential}). The actual value of the mass, used as an input to the trained network, can be computed from the different components in said mass terms. In case of the fermionic DM component, the calculation is rather simple, as it only involves $m_{\chi}$, $g_S$ and the mediator scalar field, so there is little difference to be expected between the input parameter $m_{\chi}$ and the actual value. In case of the scalar DM component, however, the calculation involves a multitude of additional parameters, which contribute to a significant difference between the input parameter and the observed mass. The computation of these mass values is performed by \textsc{SPheno} during generation of the parameter cards required by $\textsc{MadGraph5\_aMC@NLO}$ and the results can be consulted in Table \ref{tab:masses}}.

\begin{table}[!h]
\begin{center}
\begin{tabular}{|l|l|l|ll|}
\hline
\textbf{Signature}      & \textbf{Input mass parameter}                                                                                     & \textbf{Coupling constants}                                                                                                                                       & \multicolumn{2}{l|}{\textbf{Pole mass}}                                                                                     \\ \hline
Mono-jet                & \begin{tabular}[c]{@{}l@{}}$m_{\chi} \in [10, 500]$ GeV\\ $m_s \in [100, 485]$ GeV\end{tabular}                   & \begin{tabular}[c]{@{}l@{}}$g_S \in [7.4e^{-6}, 2.94]$\\ $\mu_{\phi SS} \in [4.2e^{-8}, 1200]$\\ $\lambda = 0.01$, $\lambda_{\phi \phi S S} = 0.02$\end{tabular} & \multicolumn{2}{l|}{\begin{tabular}[c]{@{}l@{}}$m_{\chi} \in [10, 570]$ GeV\\ $m_s \in [70, 540]$ GeV\end{tabular}}         \\ \hline
\multirow{2}{*}{Mono-$Z$} & \multirow{2}{*}{\begin{tabular}[c]{@{}l@{}}$m_{\chi} \in [100, 320]$ GeV\\ $m_s \in [115, 500]$ GeV\end{tabular}} & \multirow{2}{*}{\begin{tabular}[c]{@{}l@{}}$g_S = 0.2$\\ $\mu_{\phi SS} = 125$\\ $\lambda = 0.057$, $\lambda_{\phi \phi S S} = 0.02$\\ \end{tabular}}               & \multicolumn{1}{l|}{1DM} & \begin{tabular}[c]{@{}l@{}}$m_{\chi} \in [230, 1000]$ GeV\\ $m_s \in [80, 230]$ GeV\end{tabular} \\ \cline{4-5} 
                        &                                                                                                                   &                                                                                                                                                                   & \multicolumn{1}{l|}{2DM} & \begin{tabular}[c]{@{}l@{}}$m_{\chi} \in [200, 650]$ GeV\\ $m_s \in [90, 360]$ GeV\end{tabular}  \\ \hline
\end{tabular}
\end{center}
\caption{Values of the input mass ranges for each DM components for both studied signatures, along with the input mass parameters and the relevant coupling constants. Notice that, for the mono-$Z$ signature, the mass ranges differ between the one- and two-component DM cases.}\label{tab:masses}
\end{table}

\subsection{Kinematic Distributions}

Each of the data files, obtained by Monte Carlo (MC) event generation as previously described, contains $10,000$ simulated events for the selected processes. The events are then reconstructed, allowing for a representation of the relevant distributions (i.e., those to be fed to the NN) for both signatures, which, again, we treat in turn. 

\noindent For the mono-jet case we require the existence of one jet with pseudorapidity $|\eta|<2.5$ and veto against additional jets with $p_T>20$ GeV. For the mono-$Z$ case we require two opposite-sign same flavor leptons (either electrons or muons) with an invariant mass in the range $80<m_{ll}<100$ GeV.

\subsubsection{Mono-Jet}

As mentioned previously, the relevant distributions for the mono-jet case are the transverse momentum/energy (of the jet/missing one) and the pseudorapidity of the jet. At the level of reconstructed events the MET and jet transverse momentum are not exactly equal in magnitude (unlike at the partonic level) as there will be additional soft QCD radiation that will not be reconstructed as a jet.

\noindent As depicted in Figure~\ref{fig:monojet-single-dists}, in the case where only one DM particle is produced, e.g., the fermionic one, the slopes of both the  $p_T$ (of the jet) and MET distributions are quite useful in identifying the mass. The pseudorapidity also exhibits a slight dependence on the mass and is thus also considered in the NN analysis. As the production cross section of DM + jet can vary, we consider only the shapes of the kinematical distributions (normalize all distributions to $1$), not their absolute event rates.

\begin{figure}[!t]
    \centering
    \includegraphics[width=0.33\linewidth]{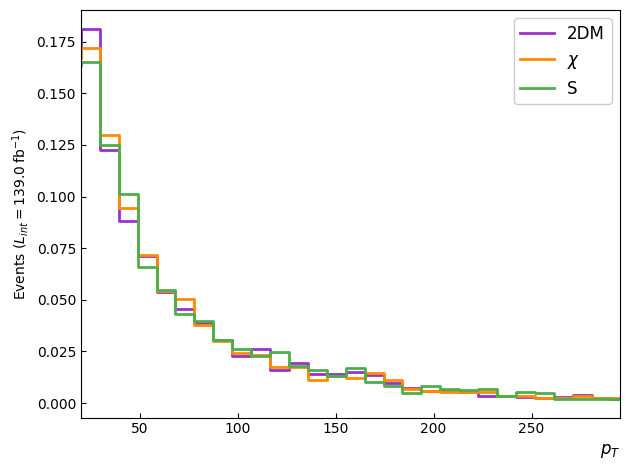}
    \includegraphics[width=0.33\linewidth]{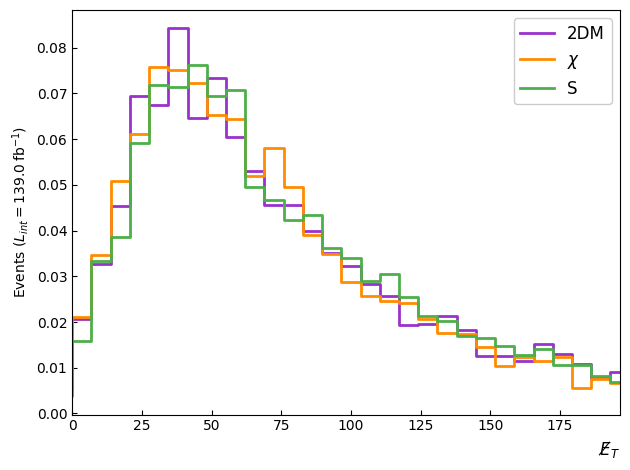}
    \includegraphics[width=0.33\linewidth]{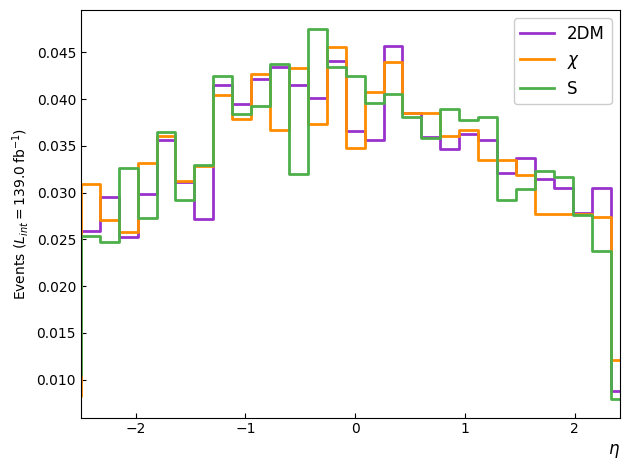}
    \caption{Studied distributions for the mono-jet signature for each of the different signals. From left to right, they are $p_T$, MET and $\eta$.}
    \label{fig:monojet-single-dists}
\end{figure}

\noindent The expected capabilities of the NN with the information given by the relevant distributions of this signature are to be able to classify the signals by the amount of DM components that they contain and then estimate their masses. We will prove this in a forthcoming section.

\subsubsection{Mono-$Z$}

The leptonic final state of this process offers the possibility to gain insight on an additional property of the DM particles with respect to the mono-jet case, i.e., the DM spin. As discussed in a previous section, this is done through $\Delta\phi$. Figure \ref{fig:monoZ-single-dists}  illustrates analogous distributions to those presented in Figure \ref{fig:monojet-single-dists}, for the three usual observables ($p_T$ (of the lepton), MET and $\eta$ of the lepton) alongside the additional one ($\Delta\phi$). 

\begin{figure}[!t]
    \centering
    \includegraphics[width=0.35\linewidth]{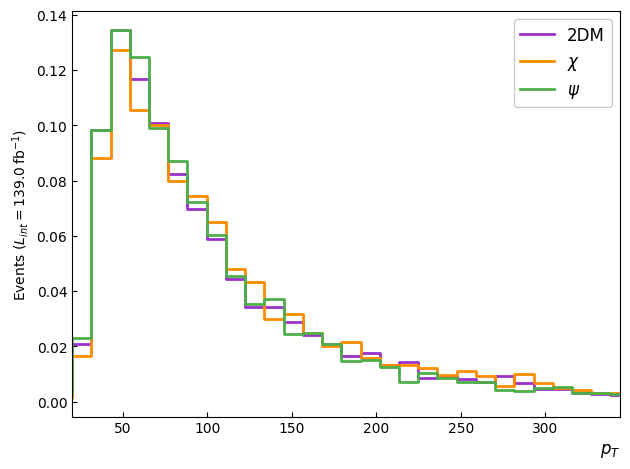}
    \includegraphics[width=0.35\linewidth]{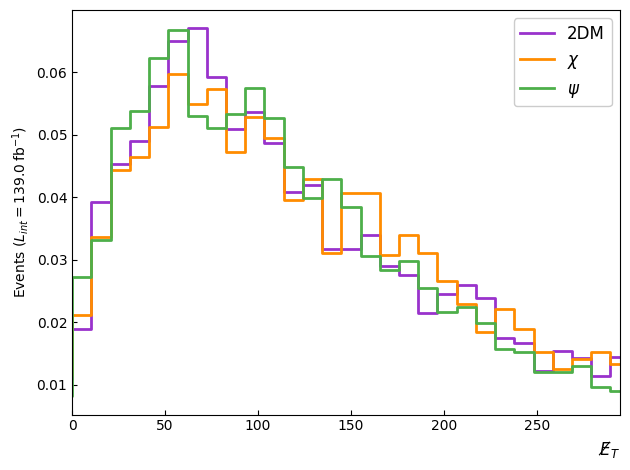}
    \includegraphics[width=0.35\linewidth]{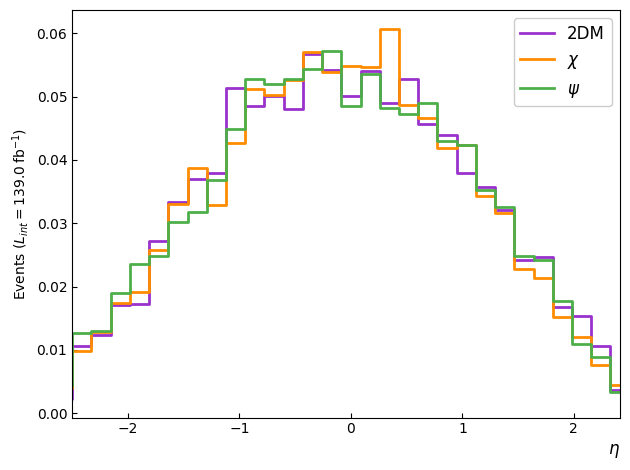}
    \includegraphics[width=0.35\linewidth]{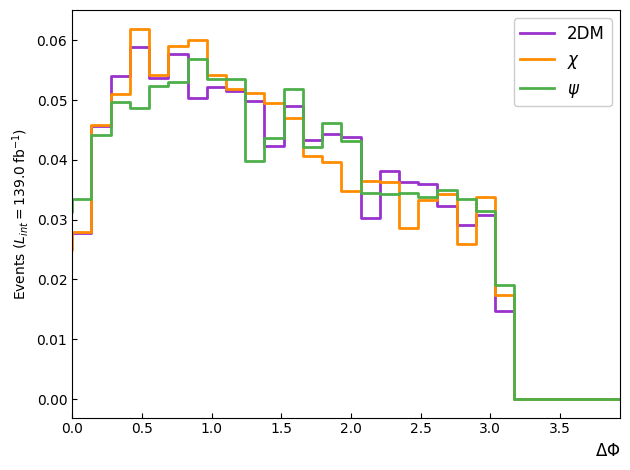}
\caption{Studied distributions for the mono-$Z$ signature in the case where only one DM particle (here, the DM fermion, $\chi$) is produced. From left to right, they are $p_T$ and MET (top) plus $\eta$ and $\Delta\phi$ (bottom).}
    \label{fig:monoZ-single-dists}
\end{figure}

\noindent In this case, the information given allows the NN to classify the different signals by the amount of DM components they contain along with a differentiation of the one-component DM samples by spin. The mono-$Z$ signal is also expected to allow for a more precise mass estimation than the mono-jet one. 

\section{Dark Matter Characterization with ML }
\label{sec:ML}

 \noindent{Here, our focus is on inferring the number of DM components, along with their spin and mass. When dealing with discrete parameters like the number of DM components and spin, a classification task is employed while a regression task is used for continuous parameters, such as mass. The workflow for our DM characterization process for mono-jet and mono-$Z$ signals is illustrated in Figures~\ref{fig:monoj_flowchart} and \ref{fig:monoz_flowchart}, respectively. While both mono-$X$ signals are used  for all purposes, it turns out that mono-jet ones are already effective in differentiating the number of DM components while mono-$Z$ ones are best to also differentiate between the spins of DM. In short, we first distinguish between the types of signals using the classifier, then we pass the input data to the corresponding regressor to infer the DM mass. The classification and the regression problems are considered separately to avoid error propagation. }
 % to explain the workflow
 
\begin{figure}[!ht]
    \centering
    \includegraphics[width=0.85\linewidth]{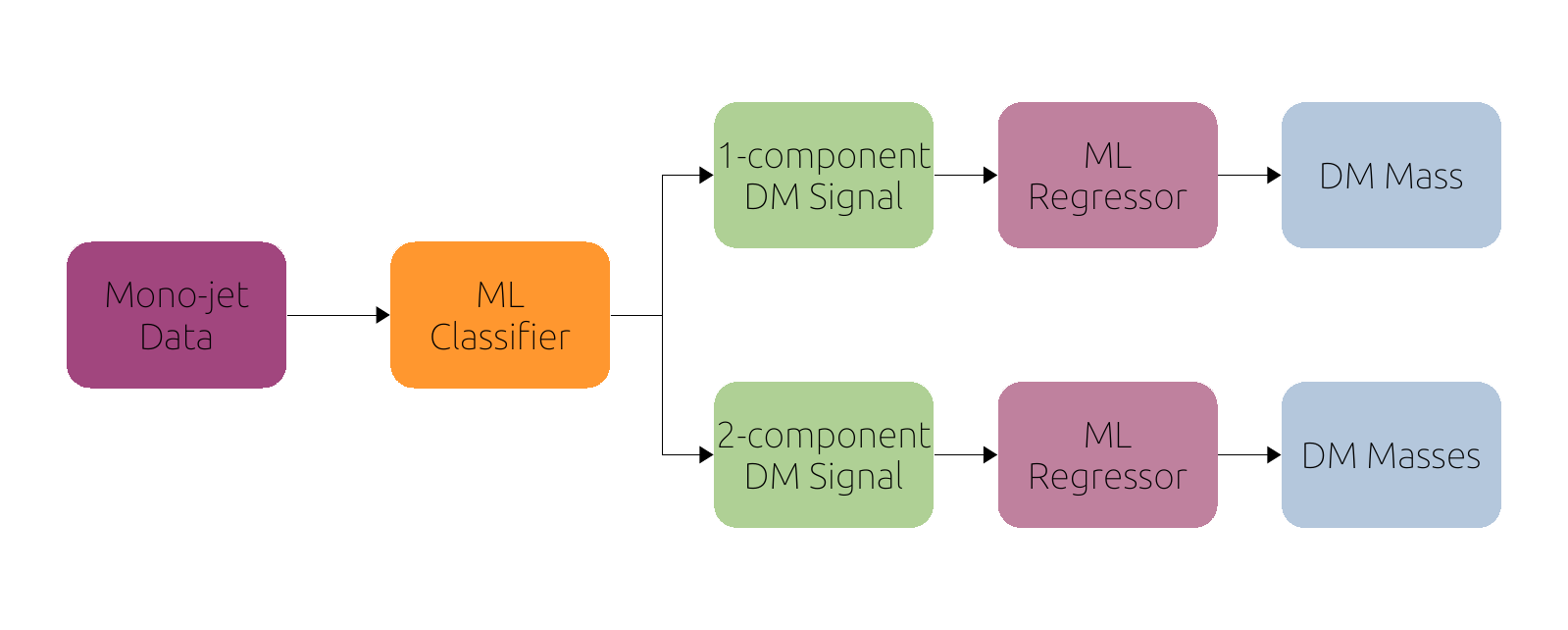}
    \caption{Visual representation of the workflow for the mono-jet signature with 1 ML classifier and 2 ML regressor. Note that the classifier only allows for a distinction between one- and two-component DM signals. }\label{fig:monoj_flowchart}
\end{figure}
\begin{figure}[!ht]
    \centering
    \includegraphics[width=0.85\linewidth]{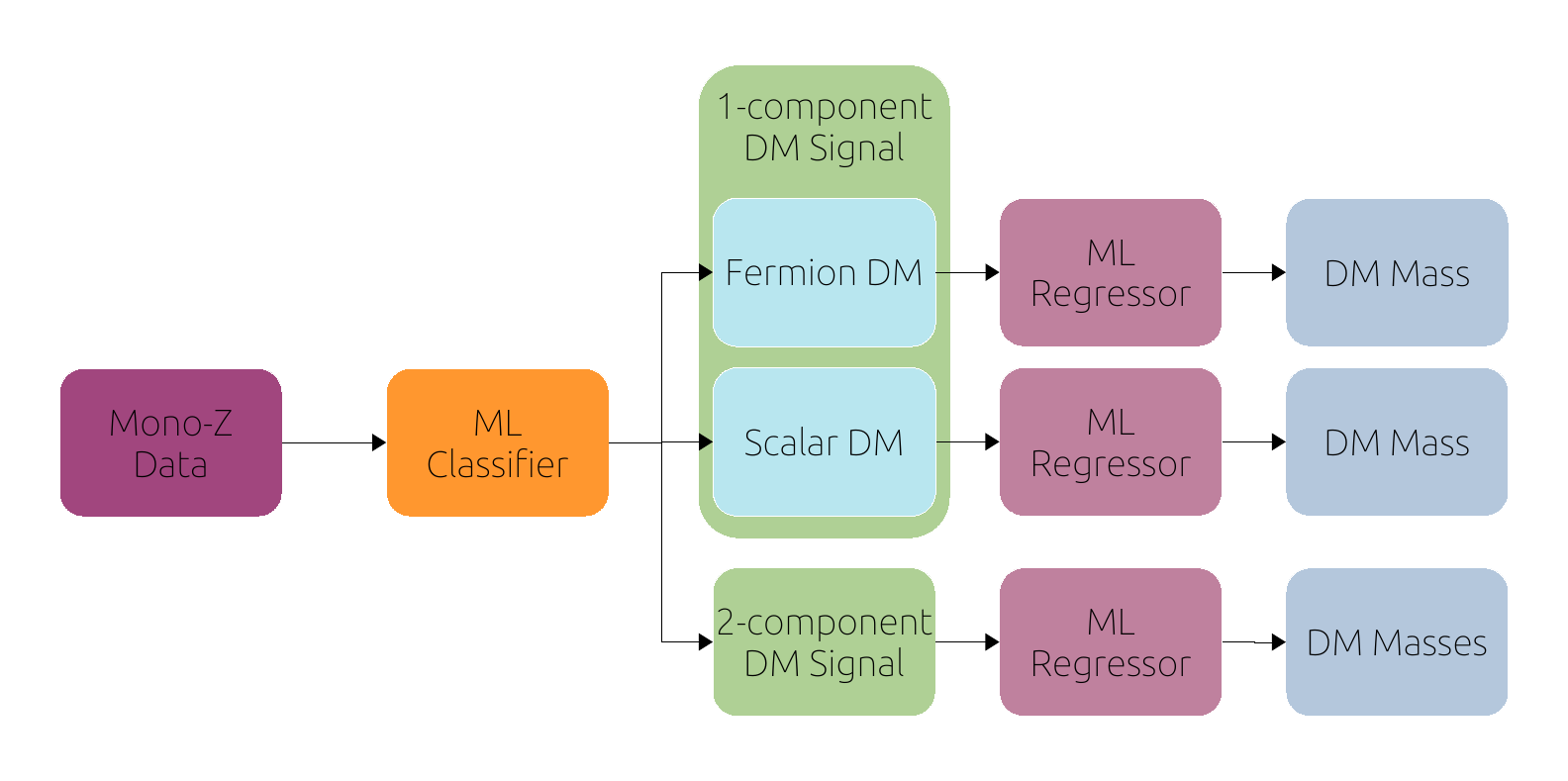}
    \caption{Visual representation of the workflow for the mono-$Z$ signature with 1 ML classifier and 3 ML regressor. Unlike for the mono-jet signature, the network is able to distinguish the one-component DM signals by spin. This is implemented in the same classifier that distinguishes the number of DM particles within a signal.} \label{fig:monoz_flowchart}
\end{figure}

\noindent  As our study is primarily a proof-of-concept about establishing the potential of mono-jet and mono-$Z$ signals in characterizing DM properties, we reinstate here that we have assumed the signal-only hypothesis with no background. That is, we implicitly acknowledge that the efficiency of any inference will depend strongly on the signal-to-background ratio \cite{Arganda:2021azw, Khosa:2019kxd}. Considering that different parameter configurations will lead to different cross sections and, subsequently, to differences in magnitude of background contamination, we expect varying performance in relation to the inferred parameter. There are of course multiple ways that one can improve the signal-to-background ratio in a realistic analysis \cite{Metodiev:2017vrx, Nachman:2020lpy, Hallin:2021wme}, but that is not the main focus of this work. In short, the results presented here can only be seen as the best-case scenario. 
 % to explain the assumptions, caveats and shortcomings of this work

\subsection{The \nobs-channel 1-Dimensional CNN (n-1DCNN)}

\noindent We opted for an \nobs-channel 1-Dimensional CNN (n-1DCNN) to be our ML classifier and regressor. Figure \ref{fig:ML-architecture} shows a schematic illustration of an n-1DCNN. Histograms for each observable are treated as a distinct channel (or "color") and the normalized bin counts are regarded as the "pixels". As the same data for each mono-$X$ signature will be used for both classification and regression task, the main differences in ML architecture between them will be the hyperparameters and loss functions used. In general, the n-1DCNN starts with 1-dimensional convolutional layers with dropout and maxpooling layers between them, followed by a flatten layer and dense layers. The specifics of each of the n-1DCNN used are presented in Sections \ref{sec:class} and \ref{sec:reg} accordingly. Here, it is important to emphasize the need for a flatten layer and dense layers after the convolutional layers. As there can be spatial correlation between the bins in certain observables (for instance $p_T$ and $\eta$), breaking our data to an n-1DCNN would disentangle the correlation between them. Thus, we need to ensure that the correlations between different observables are reintroduced via the latter layers. Maxpooling layers also play a crucial role as it allows our network to be robust towards statistical noise that the histogram may have (i.e., instead of learning specific noise patterns, it learns properly the Probability Density Function (PDF)). 
% to explain what is a n1DCNN

\begin{figure}[!ht]
    \centering
    \includegraphics[width=\linewidth]{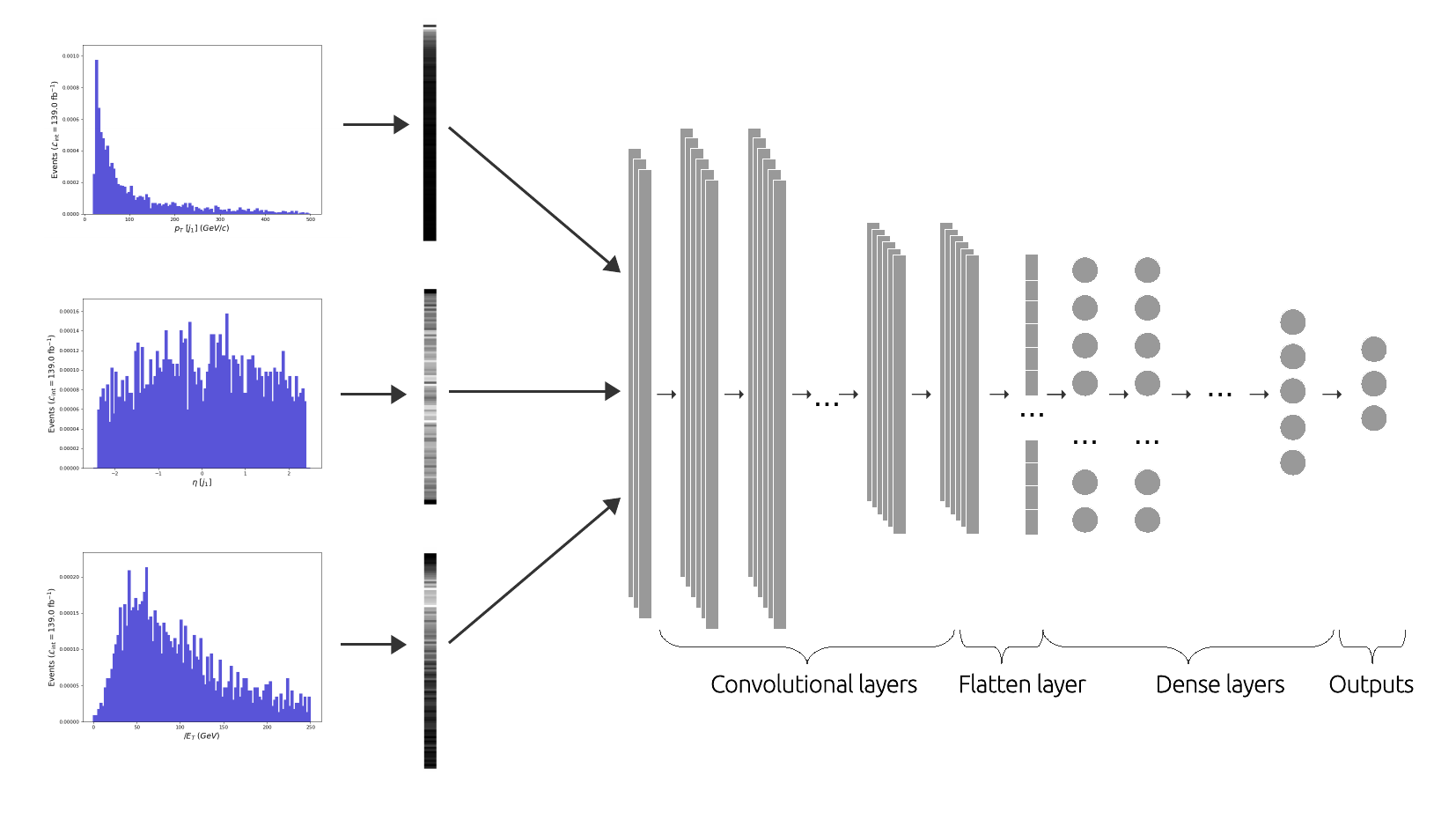}
    \caption{Schematic illustration of an n-1DCNN with 3 observables.}\label{fig:ML-architecture}
\end{figure}

\noindent One of the reasons why we employ such architecture is that it can be extended to any number of observables, \nobs. Prior studies \cite{Arganda:2021azw, Khosa:2019kxd} on DM characterization with ML had mainly focused on the transverse momentum $p_T$ and the pseudorapidity $\eta$ of the mono-jet. This allowed them to use a 2D-CNN. In our scenario, we are dealing with 3 observables (MET, $p_T$ and $\eta$) for mono-jet and 4 observables (MET, $p_T$, $\eta$ and $\Delta \phi$) for mono-$Z$, as discussed in Section \ref{sec:data}.\footnote{Considering the use case in other Beyond the SM 
(BSM) characterization that may have multiple final state, finding a suitable summary statistics to reduce the number of observables might not be trivial. An n-final state process would lead to at least 3n observables (if we take only $p_T, \eta $ and $ \phi$ of each reconstructed object)} This will also be relevant when dealing with Next-to-LO (NLO) processes as there can be additional observables (such as the $p_T$ and $\eta$ of extra jets). In \cite{Khosa:2019kxd}, Principal Component Analysis (PCA) was used to identify important observables. However, if only 2 observables were used out of the principle components, this could also potentially lead to information loss and lower efficiency. {{In such a paper, the authors did both a LO and NLO analysis. For the LO one, there was no need for PCA but, for the NLO one, they used PCA to reduce the number of observables (even then, though, only the 2 best ones were used in the ML analysis).}} An n-1DCNN also ensures that simulation-based inference beyond 2 observables can be performed via a CNN with less penalty on the number of parameters to 
train.\footnote{Recall that the introduction of CNNs was to handle the rapid increase in the number of trainable parameters in a Multi-Layer Perceptron (MLP) with the number of pixels if each pixel was an input.} Of course, one can also choose a \nobs-dimensional CNN instead. However, this will get complicated very fast as the kernel needs to parse through \nobs-dimensions, thus eventually reaching a bottleneck where even reading data will be excessively time-consuming. A n-1DCNN would then be a healthy balance between  performance and versatility. 
% to explain why we chose n1DCNN

\noindent When it comes to data representation, one could also suggest the use of event-by-event approach rather than a histogram-based approach. It was shown \cite{Nachman:2021yvi} that under the assumption that collider events are Independent and Identically Distributed (IID), per-ensemble classifier and the composite per-event classifier have the same asymptotic information content. This suggests that it is only a matter of practicality whether to use one over the other. However, this is true only if we were to ignore nuisances parameters \cite{Heinrich:2023bmt}. Stability in performances was also an issue using the event-by-event approach \cite{Arganda:2021azw}. Finally, our interest is in inferring BSM parameters that influence the entire data set. Thus, using a data set wide approach should be more suitable to capture the global characteristics of the data efficiently. 

\begin{table}[!h]
\begin{center}
\begin{tabular}{|c|c|c|}
\hline
\textbf{Observable} & \textbf{Mono-jet} & \textbf{Mono-${\bf {\it Z}}$}  \\ \hline
$p_T$               & {[}20,300{]} GeV  & {[}20,350{]} GeV \\ \hline
$\not\!\!\!\!E_T$   & {[}0, 200{]} GeV  & {[}0, 300{]} GeV \\ \hline
$\eta$              & {[}-2.5, 2.5{]}   & {[}-2.5, 2.5{]}  \\ \hline
$\Delta \Phi$       & -                 & {[}0, $\pi${]}   \\ \hline
\end{tabular}
\end{center}
\caption{Observables utilized in the training for both mono-jet and mono-$Z$ signals. Here, $\Delta \Phi$ is defined between the two outcoming leptons in the mono-$Z$ channel and is used to determine the spin of one-component DM events, allowing for differentiation between fermionic and scalar DM.}\label{tab:observables}
\end{table}

\noindent Histograms are obtained after the cuts defined in Section~\ref{sec:data}; their details are summarized in Table~\ref{tab:observables}. All NNs were constructed using \texttt{Keras} and \texttt{TensorFlow}. Optimizations were performed using the Adaptive Moment (Adam) Optimizer with a learning rate {{between $10e^{-3}$ and $10e^{-4}$.}} The NNs were trained with a maximum of 1000 epochs. Early stopping with varying patience values were employed to avoid overfitting and the weights that gave the best performance were saved. A cross-entropy loss function was used for the classification task, while the regression task used mean squared error. {{The overall architecture of the NN is very similar for both tasks at hand and both studied signatures, as can be seen in Tables \ref{tab:NN_arch_class} and \ref{tab:NN_arch_reg}, with the main differences residing in the hyperparameters, optimized for each individual case.}}

\begin{table}[!h]
\begin{center}
\begin{tabular}{|l|c|l|}
\hline
\multicolumn{1}{|c|}{\textbf{Layer structure}}                          & \textbf{Activation function} & \multicolumn{1}{c|}{\textbf{Use}}                                                                   \\ \hline
\begin{tabular}[c]{@{}l@{}}Conv1D\\ Dropout\\ MaxPooling1D\end{tabular} & Relu                         & Feature extraction                                                                                  \\ \hline
Flatten                                                                 & -                            & \begin{tabular}[c]{@{}l@{}}Conversion of multi-dimensional\\ feature map into 1D array\end{tabular} \\ \hline
\begin{tabular}[c]{@{}l@{}}Dense\\ Dropout\end{tabular}                 & Relu                         & Classification                                                                                      \\ \hline
Dense                                                                   & Softmax                      & Output                                                                                              \\ \hline
\end{tabular}
\end{center}
\caption{General architecture of the NN utilized for the classification task. In the mono-jet case, three copies of the layer structure used for feature extraction are employed, whilst only two are used for mono-$Z$. A higher number of filters is used in the latter as a trade-off.}\label{tab:NN_arch_class}
\end{table}

\begin{table}[!h]
\begin{center}
\begin{tabular}{|l|c|l|}
\hline
\multicolumn{1}{|c|}{\textbf{Layer structure}}                                                 & \textbf{Activation function} & \multicolumn{1}{c|}{\textbf{Use}}                                                                   \\ \hline
\begin{tabular}[c]{@{}l@{}}Conv1D\\ Conv1D\\ MaxPooling1D\\ Conv1D\\ MaxPooling1D\end{tabular} & Relu                         & Feature extraction                                                                                  \\ \hline
Flatten                                                                                        & -                            & \begin{tabular}[c]{@{}l@{}}Conversion of multi-dimensional\\ feature map into 1D array\end{tabular} \\ \hline
Dense                                                                                          & Relu                         & Regression                                                                                          \\ \hline
Dense                                                                                          & -                            & Output                                                                                              \\ \hline
\end{tabular}
\end{center}
\caption{General architecture of the NN utilized for the regression task. Here there is no difference in the structures between signatures nor between single and multi DM component cases, the only difference thus residing in the values of the hyperparameters themselves.}\label{tab:NN_arch_reg}
\end{table}

\noindent {{For each of the one-component DM scenarios we generated 600 mass points, whilst 1,200 were generated for the two-component DM ones, with the aim of having the same amount of available events for both multiplicities. This amounts to a total 2,400 mass points for each signature.\footnote{These numbers refer solely to the main case of study of this paper, where the cross sections of both the fermionic and scalar DM component are similar (as given by equation (\ref{eq:XS_ratios})). Considering the other studied cases, where $r$ is $2$, $3$ and $5$, for which we generated the same amount of mass points, the total number goes up to 9,600 points for each signature.} The data set for each task is split into the three usual subgroups: training, validation and testing, with a ratio of $0.64 : 0.16 : 0.2$, respectively.}}

\noindent {{In order to increase the statistical significance of the results, each task is performed $50$ separate times, utilizing the same mass points but a different split seed.}}

\subsection{Classification: Number of DM Components and DM Spin} \label{sec:class}

{For both signatures, the NN is built to classify the signal as coming from either one-component DM fermion, one-component DM scalar or two-component DM, utilizing the aforementioned distributions. When dealing with mono-$Z$ events, it is expected for the network to be able to distinguish between the spins of the particles in the one-component DM  case, given the angular distance between the two leptons in the final state. This feat is not possible in the mono-jet scenario, given the lack of information available in the given observables. This signature is, therefore, only expected to be able to distinguish between single and two component DM.

\begin{figure}[!ht]
    \centering
    \includegraphics[width=0.45\linewidth]{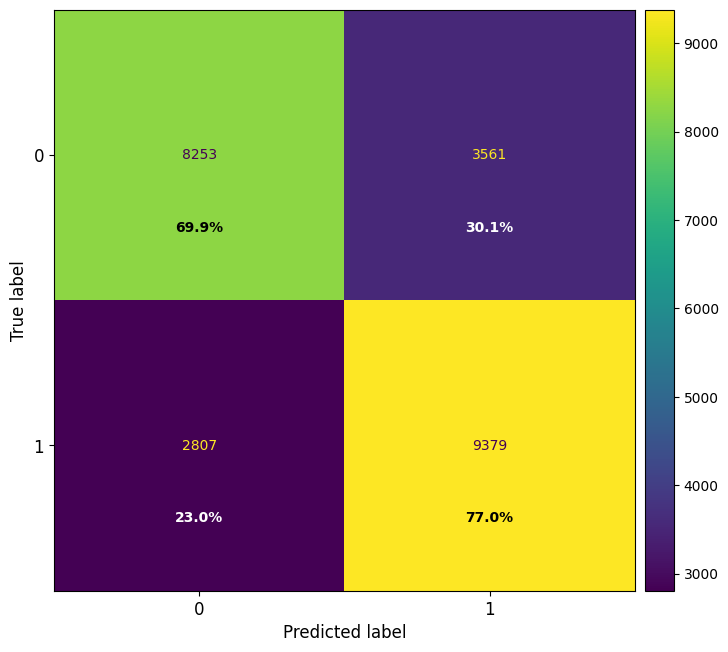}
    \includegraphics[width=0.45\linewidth]{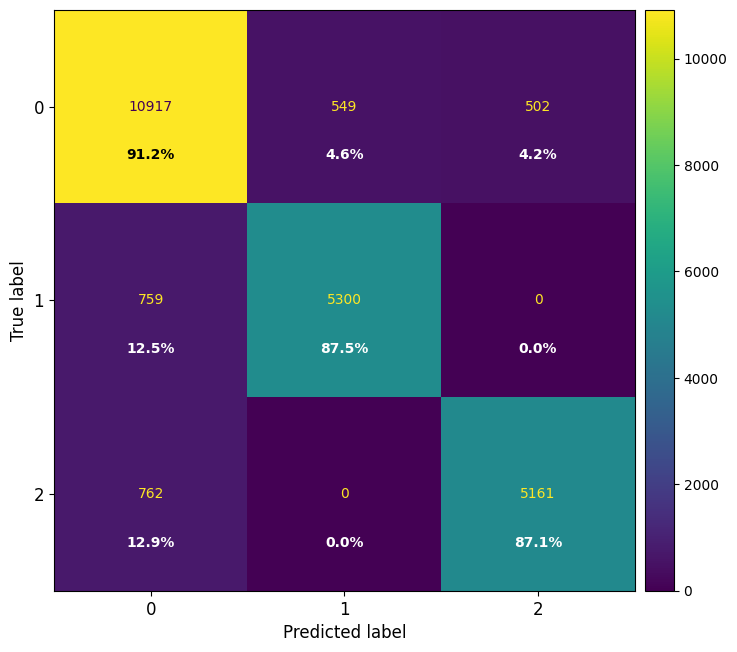}
    \caption{Confusion matrix for mono-jet (left) and mono-$Z$ (right). The results for the testing of all independent runs are shown here. The percentages in each 'true label' add up to $100$, meaning that, e.g., out of all tested samples with true label $0$ (two-component DM) in the mono-jet signature, nearly $70\%$ of them were correctly tagged, while the remaining $30\%$ was instead tagged as being one-component DM ($1$).}\label{fig:cm}
\end{figure}

\noindent {The performance of the network can also be studied through the Receiver Operating Characteristic (ROC) curves, displayed in Figure \ref{fig:roc}. The achieved AUC in both signals match the results observed in the confusion matrices.}
 
\begin{figure}[!ht]
    \centering
    \includegraphics[width=0.45\linewidth]{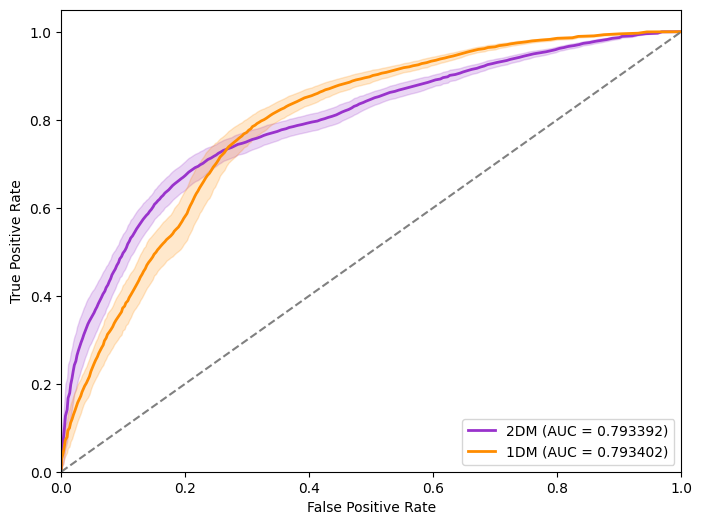}
    \includegraphics[width=0.45\linewidth]{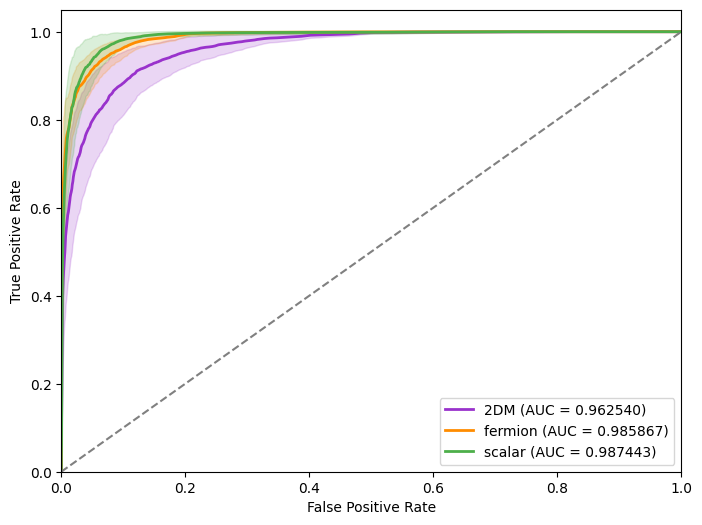}
    \caption{ROC curve for NN classification of DM signals for the mono-jet data (left) and mono-$Z$ data (right). The plotted curve is the average of the aforementioned $50$ independent runs, with the semi-transparent bands showing the spread of said runs.}\label{fig:roc}
\end{figure}

\subsection{Regression: DM Mass} \label{sec:reg}

As this study primarily serves as a proof-of-concept to demonstrate the application of 1D-CNN to infer DM properties, $\theta$ from collider data, $x_{\text{data}}$, we will only obtain point estimates $\hat{\theta}(x_{\text{data}})$ rather than other more comprehensive statistical objects such as posteriors $p(\theta | x_{\text{data}})$, likelihood ratios $p(x_{\text{data}}|\theta)/p(x_{\text{data}}|\theta_{0})$, etc.  Since we can always build upon our current work with existing methods \cite{Papamakarios:2019fms,Papamakarios:2017tec, Brehmer:2018hga, Heinrich:2023bmt}, we leave a more thorough treatment for future works.

\noindent {{Once the number of DM components in the sample has been determined using the classifier, we want to characterize it by determining its mass. We do this with the implementation of a regression task.}  The regression problem is more complex than the classification one, as the NN is expected to determine a mass (or masses) that can vary within a wide range of values. This does not affect the network structure in a noticeable manner, however, as can be seen in Tables ~\ref{tab:NN_arch_class}} and \ref{tab:NN_arch_reg}.

\noindent {{The performance of the network in the regression task is studied through three different figures of merit. First, the relative error, which can be defined by:}}

\begin{equation}
        \epsilon = \frac{m_{\rm pred}-m_{\rm true}}{m_{\rm true}}. \label{eq:rel_error}
\end{equation}

\noindent {{This allows us to visualize the distribution of the prediction deviation relative to the true value of the mass, given that a deviation of, e.g., $50$ GeV in the prediction is much more significant for  $m_{\rm true}=50$ GeV than it is for $m_{\rm true}=250$ GeV. Alongside this figure we also present the mean, $\mu$, and the standard deviation, $\sigma$, for each of the DM components (obtained over 50 runs).}

\noindent {{It is relevant to take a detailed look at the standard deviation. For that, we introduce the second figure of merit, which is $\sigma$ as a function of the true mass, allowing us to determine if our network struggles with a particular sector of the mass spectrum. Lastly, we also use the tolerance, which shows what fraction of the total number of events have an absolute mass deviation, $\Delta m$, below a certain threshold $\delta$. The absolute mass deviation is defined as: }}

\begin{equation}
        \Delta m = |m_{\rm pred}-m_{\rm true}|. \label{eq:AMD}
\end{equation}

%\textbf{Note that for mono-jet signatures, the network stops improving after the number of filters in the \texttt{Conv1D} layers is doubled (one-component signals) or quadrupled (two-component signals) with respect to the classification architecture, which means that there is no additional information left to be learnt from the data. For the mono-$Z$ signature, however, the network keeps improving until a further limit is reached, showing the difference in available information offered by only one additional kinematic distribution.} \textcolor{blue}{(YS:Why double or quadruple instead of a steady increase in filters. Also, was this the largest one before it saturates? also, is there proof to show this statement? If yes, show in a plot or something)} \textcolor{green}{(SOL: Can be easily checked by doing a for loop over number filters. Need to see the rebinned data first.} \textcolor{orange}{MFC: Did not dive further into this. Just optimized the network and got what I got. The results are very good, imo, and I've also think I've done more than my share of the work so I will not be looking into this ever. If it does not convince you, just remove this part of the discussion.}

\subsubsection{The One-component DM Signals}
\subsubsection*{I. Mono-jet}

Figure~\ref{fig:epsilon_mono-jet_1DM} shows the mass prediction error in the one-component DM case for the mono-jet signature. Despite the classifier not being able to distinguish between the fermionic and scalar DM cases, we show them separated at truth level for clarity and to investigate if there is any significant differences between the two. Given that the network sees them as the same and that the available mass range of the scalar DM is smaller than and contained within the fermionic DM one, it is expected there be a certain region of the spectrum, namely  between $[10,100]$, that has significantly fewer events and is therefore prone to more significant errors. This is immediately noticeable in the distributions in Figure~\ref{fig:epsilon_mono-jet_1DM}, having the one for the fermionic DM a large tail on the left hand side, along with the variations in the standard deviation of the relative error presented in Figure~\ref{fig:sigma_mono-jet_1DM}. 

\noindent {{The scalar DM component, in contrast, is more promising. It displays almost no predictive bias, unlike the slight tendency to under-predict the mass of the fermionic DM component, and is able to make predictions with $20\%$ resolution.}}

\begin{figure}[!ht]
    \centering
    \includegraphics[width=0.5\linewidth]{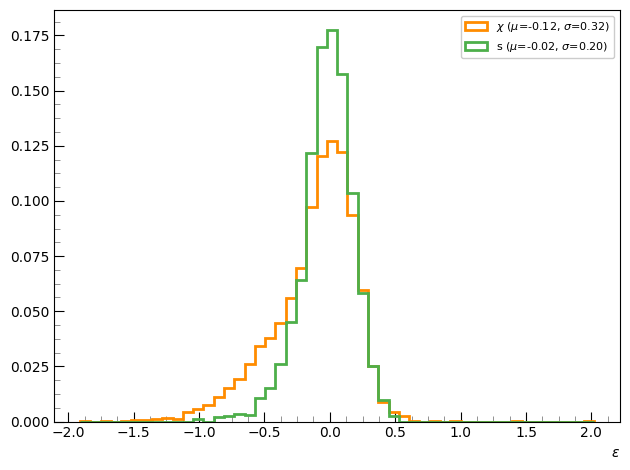}
    \caption{Relative error in the mass prediction for both DM particles in the one-component DM case for the mono-jet signature.}\label{fig:epsilon_mono-jet_1DM}
\end{figure}

\noindent {{Taking a look at Figure \ref{fig:sigma_mono-jet_1DM}, we can see that the network struggles to predict masses below $300$ GeV. There is a clear mass dependence in the predictions of the fermionic DM component, whilst the predictions of the scalar are more stable. There is, however, a noticeable increase in the performance for high masses, with the error dropping to less than a $15\%$ in both cases.}}

\begin{figure}[!ht]
    \centering
    \includegraphics[width=0.5\linewidth]{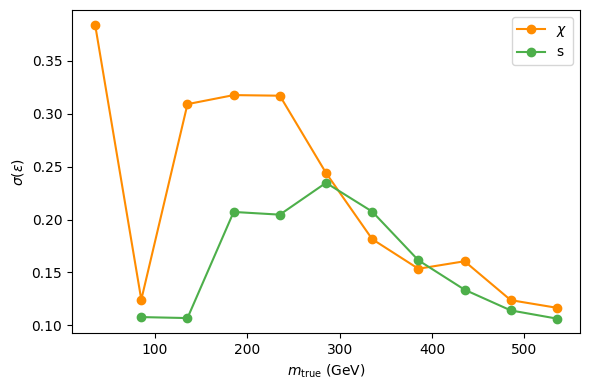}
    \caption{Standard deviation of the relative error, $\sigma(\epsilon)$, as a function of the true mass for the single component DM in the mono-jet signature. There is a significant drop in the deviation toward the high mass end of the spectrum.}\label{fig:sigma_mono-jet_1DM}
\end{figure}

\noindent {{The actual predictive power of the network is better understood by looking at Figure \ref{fig:tolerance_mono-jet_1DM}. For the scalar DM component, the network predicts the mass within $50$ GeV for $68\%$ of the tested events and within less than $150$ GeV for $90\%$ of them. As expected given the discussion, the results for the fermionic DM component are slightly worse, with the network being able to predict the mass within roughly $80$ GeV for $68\%$ of the events and within $175$ GeV for $90\%$ of them.}}

\begin{figure}[!ht]
    \centering
    \includegraphics[width=0.5\linewidth]{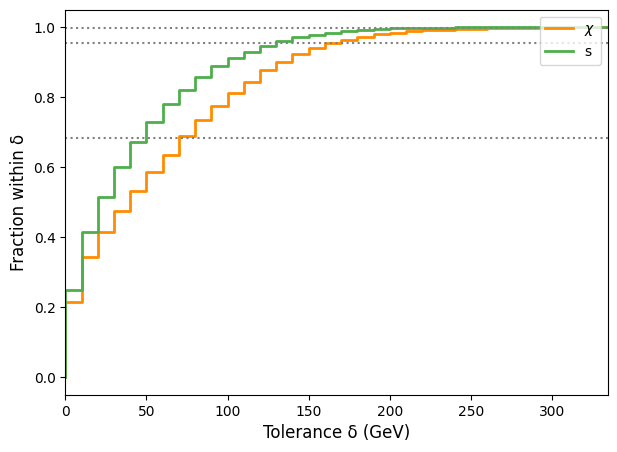}
    \caption{Fraction of events with an absolute mass deviation below a certain threshold in the one-component DM case for the
mono-jet signature. The dotted lines represent the equivalent to a $1 \sigma$, $2 \sigma$ and $3 \sigma$ interval.}\label{fig:tolerance_mono-jet_1DM}
\end{figure}

\subsubsection*{II. Mono-$Z$}

When working with the mono-$Z$ signature, the kinematic distributions can be taken from either a lepton with a fixed charge (lepton/antilepton) or by using ordering in $p_{T}$ (leading/subleading). The two distributions, {e.g.}, one for $p_{T}(e^{-})$ and one for $p_{T}(e_{\mathrm{leading}})$ will be different. Testing of both scenarios revealed a better discriminating power when utilizing a fixed charge. Hence we shall consider results based on the transverse momentum of the lepton of a given charge in the following.

\noindent {The relative error in the mass prediction, presented in Figure \ref{fig:epsilon_mono-Z_1DM}, shows a much more promising result than the one obtained for the mono-jet signature. The network displays no bias whatsoever for either of the DM components, along with a very small standard deviation. This directly affects the predictive power of the network, as can be seen in Figure \ref{fig:tolerance_mono-Z_1DM}. For the scalar DM component, the network predicts the mass with no deviation for nearly $85\%$ of the tested events and within less than $25$ GeV for the totality of these. The results for the fermionic DM component are slightly worse, with the network being able to predict the mass within less than $50$ GeV for $68\%$ of the events and within $100$ GeV for $90\%$ of them. The main source of this difference is likely the significant difference in mass ranges, as shown in Table \ref{tab:masses}}.

\begin{figure}[!ht]
    \centering
    \includegraphics[width=0.5\linewidth]{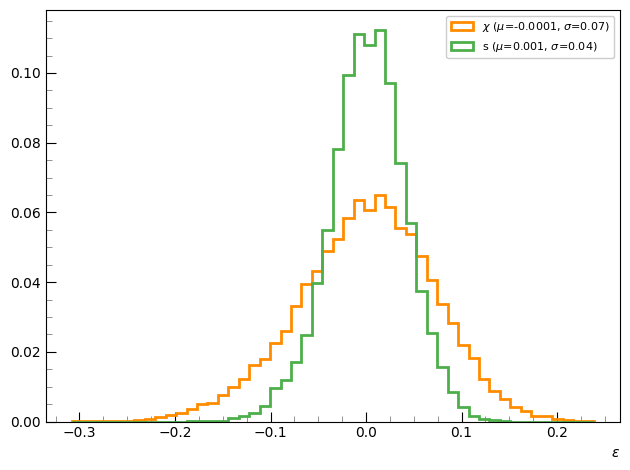}
    \caption{Relative error in the mass prediction for both DM particles in the one-component DM case for the mono-$Z$ signature.}\label{fig:epsilon_mono-Z_1DM}
\end{figure}

\begin{figure}[!ht]
    \centering
    \includegraphics[width=0.5\linewidth]{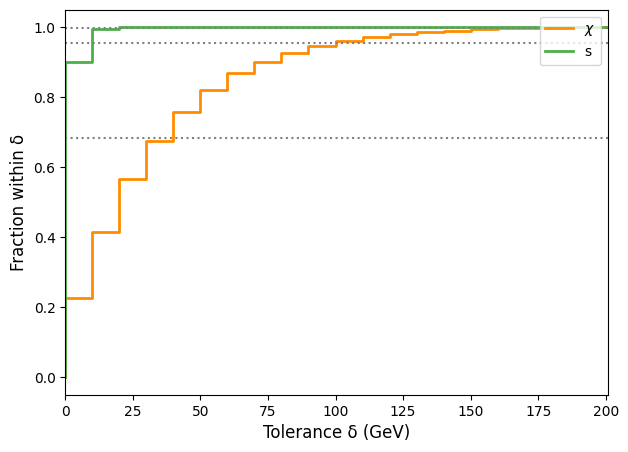}
    \caption{Fraction of events with an absolute mass deviation below a certain threshold in the one-component DM case for the
mono-$Z$ signature. The dotted lines represent the equivalent to a $1 \sigma$, $2 \sigma$ and $3 \sigma$ interval.}\label{fig:tolerance_mono-Z_1DM}
\end{figure}

\subsubsection{The Two-component DM Signals}
\subsubsection*{I. Mono-jet}

{The regression task in the two-component DM case is particularly sensitive to differences in the cross sections of the two DM components, as a consistently significant difference could result in the suppression of the component with  smaller cross section, which would in turn significantly decrease the ability of the CNN to characterize its mass. In Figure \ref{fig:epsilon_mono-jet_2DM} we can observe that this is not the case and that we are able to predict the mass with an acceptable error overall. It is important to note, however, that both the mean and the standard deviation are significantly higher here than they were for the one-component DM case. This is somewhat expected given each event contains both DM components with a very similar cross section.}

\noindent {Additionally, notice that the tail observed on the left hand side of the fermionic DM component distribution in Figure \ref{fig:epsilon_mono-jet_1DM}
is still appearing here, albeit somewhat suppressed, and now is present for both components. (The distribution extends on the right hand side solely to capture a few statistically insignificant points reaching $\epsilon=8$ or so.)}

\begin{figure}[!ht]
    \centering
    \includegraphics[width=0.5\linewidth]{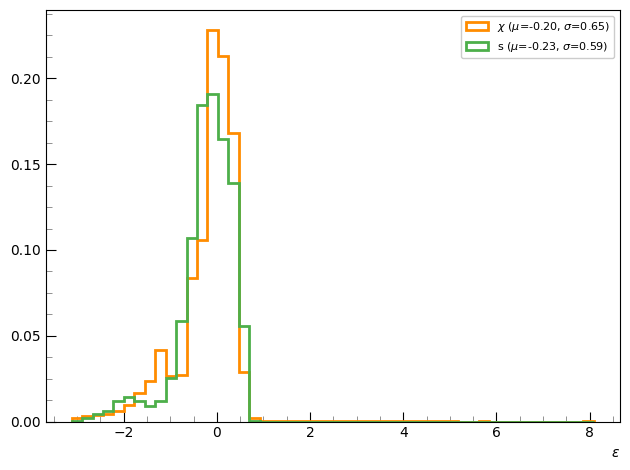}
    \caption{Relative error in the mass prediction for both DM particles in the two-component DM case for the mono-jet signature.}\label{fig:epsilon_mono-jet_2DM}
\end{figure}

\noindent {{Figure \ref{fig:sigma_mono-jet_2DM} easily explains the values of the standard deviation given by Figure \ref{fig:epsilon_mono-jet_2DM}. As mentioned for the 
one-component DM case, this is due to there being significantly more events above $100$ GeV. Here, given the added difficulty in the mass determination due to the presence of both DM components in the samples at once, the predictive power of the NN in that particular interval suffers much more. For mass values above $100$ GeV, however, we observe a rather constant deviation between $15\%$ and $30\%$ for both the fermionic and the scalar DM, which is rather similar to what we observed in the 
one-component DM case.}}

\begin{figure}[!ht]
    \centering
    \includegraphics[width=0.5\linewidth]{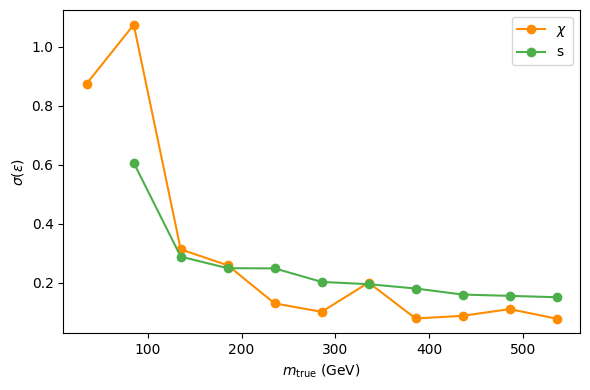}
    \caption{Standard deviation of the relative error $\sigma(\epsilon)$, as a function of the true mass in the two-component DM case for the mono-jet signature. With the exception of the lower mass range, where there are less points, the distribution is rather stable within all the mass range.}\label{fig:sigma_mono-jet_2DM}
\end{figure}

\noindent {{The discussed difficulties in the determination of the mass of both DM components at the same time are perfectly displayed in Figure \ref{fig:tolerance_mono-jet_2DM}, where a significant loss in predictive power can be noticed when comparing it to the one-component DM case. Here, the network is able to predict the mass within roughly $100$ GeV for $68\%$ of the tested events, and within less than $250$ GeV for $90\%$ of them.}}

\begin{figure}[!ht]
    \centering
    \includegraphics[width=0.5\linewidth]{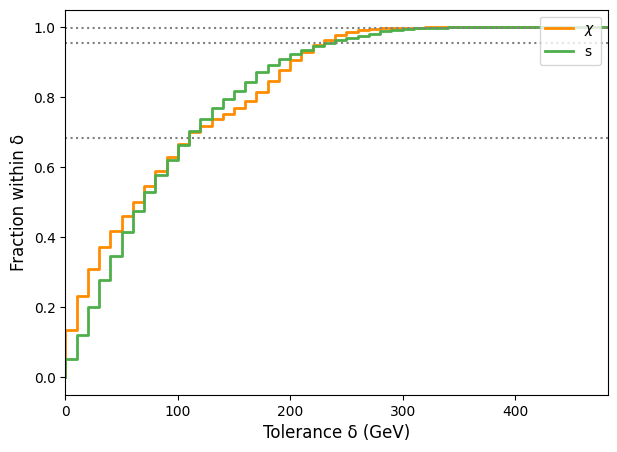}
    \caption{Fraction of events with an absolute mass deviation below a certain threshold in the two-component DM case for the mono-jet signature. The dotted lines represent the equivalent to a $1 \sigma$, $2 \sigma$ and $3 \sigma$ interval.}\label{fig:tolerance_mono-jet_2DM}
\end{figure}

\noindent {The noticeable  difference in predictive power between the one- and two-component DM networks brings the difficulty of the task at hand to the forefront. This of course may very well be caused by the similarity between the cross sections of both DM components.} %Studies in Section {AAAAAAAAA} show that there may be a certain ratio between the cross sections that allows for a better reconstruction of both masses.} 

\subsubsection*{II. Mono-$Z$}

{Much like it was observed for the mono-jet signature, the mono-$Z$ one also displays a drop in the predictive power of the network in the two-component DM case. The network exhibits a clear bias to predict mass values significantly lower than their true value, as seen in Figure \ref{fig:epsilon_mono-Z_2DM}. This, together with the much larger standard deviation, comparable to the one observed for the mono-jet signature, makes for a less reliable mass predictor.} 

\begin{figure}[!ht]
    \centering
    \includegraphics[width=0.5\linewidth]{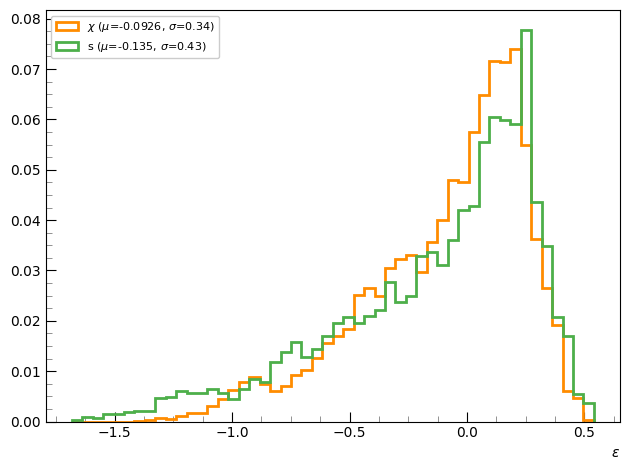}
    \caption{Relative error in the mass prediction for both DM particles in the two-component DM case for the mono-$Z$ signature.}\label{fig:epsilon_mono-Z_2DM}
\end{figure}

\noindent {The standard deviation is rather constant for mass values $m_s\gtrsim150$ GeV and $m_{\chi}\gtrsim 300$ but varies significantly below these. This directly affects the predictive power of the network, which is of $75$ GeV for $68\%$ of the tested events for the scalar DM component and $125$ GeV for $68\%$ of the fermionic DM one, roughly coinciding with the $90\%$ mark for the scalar. In the case of the fermionic DM, the $90\%$ mark is at nearly 
$225$ GeV.}

\begin{figure}[!ht]
    \centering
    \includegraphics[width=0.5\linewidth]{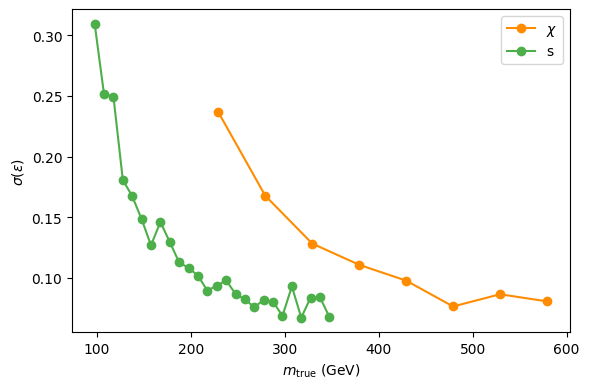}
    \caption{Standard deviation of the relative error, $\sigma(\epsilon)$ as a function of the true mass in the two-component DM case for the mono-$Z$ signature. With the exception of the lower mass range, where there are less points, the distribution is rather stable within the whole mass range.}\label{fig:sigma_mono-Z_2DM}
\end{figure}

\begin{figure}[!ht]
    \centering
    \includegraphics[width=0.5\linewidth]{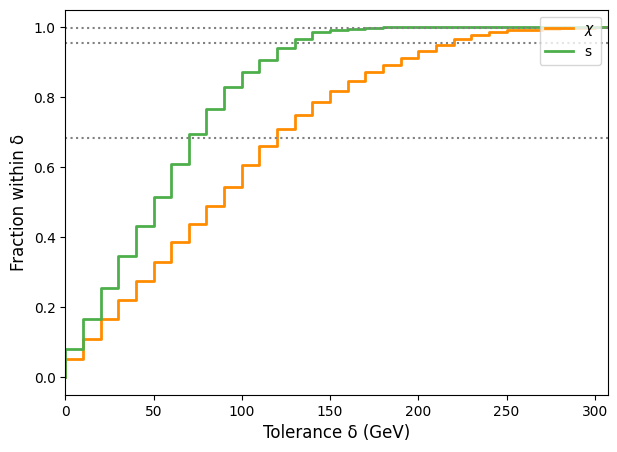}
    \caption{Fraction of events with an absolute mass deviation below a certain threshold in the two-component DM case for the mono-$Z$ signature. The dotted lines represent the equivalent to a $1 \sigma$, $2 \sigma$ and $3 \sigma$ interval.}\label{fig:tolerance_mono-Z_2DM}
\end{figure}

\newpage

\section{Conclusions}

In this article we have implemented a CNN to identify and characterize one- and two-component DM signals with information extracted from two different mono-$X$ signals, namely mono-jet and mono-$Z$. For the mono-jet signature, the network was expected to distinguish the signals by number of DM components and then determine their masses. For the mono-$Z$, the network was expected to distinguish the signals by the number of components and then identify if a one-component DM signal corresponds to the fermion or the scalar in addition to giving an estimate of the masses. (Here, the sensitivity to the spin arises from the possibility of having final state radiation, {i.e.}, $Z$ bosons emitted from the DM particles themselves.)
The results show that the CNN is able to perform the classification task with high accuracy. The regression task is notoriously more complex, which is here compensated through the implementation of more filters in the convolutional layers. 

\noindent In one-component DM signals from the mono-jet signature, the network made predictions {within $50$ GeV for $68\%$ of the tested events and within less than $150$ GeV for $90\%$ of them for the scalar DM component, with the results for the fermionic DM component being only slightly worse. For the two-component DM case, the predictive power of the network suffered notably, showing a deviation of up to $250$ GeV for $90\%$ of the events. The network presented a rather severe bias in this case, tending to predict masses lower than their true value, along with a very big standard deviation for low mass values, that decreased to somewhere between $15\%$ and $30\%$ for higher masses. This bias was not as striking for the one-component DM case and the variations in the deviations followed a similar pattern, albeit being smaller in size.}
\noindent {Altogether, the predictive power of the CNN in the mono-jet signature was consistently worse for the fermionic DM component, which may have been caused by a less populated low mass region.}

\noindent The mono-$Z$ signature assumes that a $Z$ boson is produced along with the DM pair and then decays into a pair of leptons (electrons and muons). There was clearly an incentive to check whether the network shows a bias for either of the leptons, but the results illustrated that this does not seem to be the case. For both of them, the predictions on the masses of the components of the signal improve significantly, especially on the biggest prediction error. When considering two-component DM signals, this signature also struggles to characterize one of the components, in this case $\chi$. For this signature the cross sections of both components were of the same order of magnitude and with differences in ratio of not more than $2$ units. It is possible that generating more data would help the characterization of the particle with the consistently smaller cross section, but it is also possible that it is a persistent problem.
\noindent Here, the network struggled to characterize the fermion in the lower half of the mass range when considering the kinematic distributions of the lepton while for the anti-lepton the conflict points are scattered throughout the whole mass range. For the scalar, however, the conflict points are located in the upper section of the mass range for both the lepton and anti-lepton. 

\noindent To sum up, the network is able to distinguish DM signals by their number of components and, if in the dataset there is information on the spin, it is also able to identify  one-component DM signals in it. The regression task, mainly consistent on the determination of the mass of the DM particles present in the signal, is performed with better accuracy when using mono-$Z$ signatures wherein one can exploit more kinematic observables, given the additional degrees of freedom of the final state (owing to the $Z$ decaying).

\noindent While our work clearly does not boast any claim of definiteness when coming to real-world analyses, given that we have not discussed the presence of any background, neither in mono-jet nor in mono-$Z$ signals, we believe to have opened an interesting path, based on established ML methods, to pursue the characterization of would-be DM signals at the (HL-)LHC, whether one- or two-component, a task that we would very much hope to be soon on the agenda of the CERN collider, owing to a discovery of such a new state of matter.
\section*{Acknowledgments}
SM is supported in part through the NExT Institute and the STFC Consolidated Grant ST/X000583/1.  We all thank Harri Waltari, Prashant Singh and Samuel Ramos for illuminating discussions. We further thank  Harri Waltari for actual collaboration in the initial stages of this research. Computational resources were provided by the National Academic Infrastructure for Supercomputing in Sweden (NAISS), funded by the Swedish Research Council under the project NAISS 2023/22-305, NAISS 2023/22-498, NAISS 2023/23-411, NAISS 2024/22-277, NAISS 2024/23-168, NAISS 2025/23-201, NAISS 2025/22-512 and NAISS 2026/4-806. The computations and data handling were enabled by resources provided by Chalmers e-Commons at Chalmers, PDC Center for High Performance Computing at KTH Royal Institute of Technology and UPPMAX at Uppsala University.

\end{document}